\documentclass[aps,prd,floatfix,nofootinbib,showpacs,amsfonts,superscriptaddress]{revtex4}
\usepackage{hyperref}
\usepackage{amssymb}
\usepackage{amsmath,color}
\usepackage{epsfig}
\usepackage{graphicx,epsf,epsfig}
\usepackage{footnote}
\usepackage{ulem}
\usepackage{color}

\newcommand{\sm}{{standard model }}
\def\nn{\nonumber}

\allowdisplaybreaks \allowdisplaybreaks[2]
\newcommand{\AddrDmp}{Department of Modern Physics, University of Science and Technology of China\\
  Hefei, Anhui 230026, CHINA}

\newcommand{\AddrAHEP}{AHEP Group, Institut de F\'{i}sica Corpuscular --
  C.S.I.C./Universitat de Val\`{e}ncia, Parc Cientific de Paterna.\\
  C/Catedratico Jos\'e Beltr\'an, 2 E-46980 Paterna (Val\`{e}ncia) - SPAIN}

\begin{document}
\title{Warped flavor symmetry predictions for neutrino physics}

\author{Peng Chen}
\email{pche@mail.ustc.edu.cn}
\affiliation{\AddrDmp}
\author{Gui-Jun Ding}
\email{dinggj@ustc.edu.cn}
\affiliation{\AddrDmp}
\author{Alma. D. Rojas}
\email{alma.rojas@ific.uv.es}
\affiliation{\AddrAHEP}
\author{C. A. Vaquera-Araujo}
\email{vaquera@ific.uv.es}
\affiliation{\AddrAHEP}
\author{J. W. F. Valle}
\email{valle@ific.uv.es}
\homepage[URL:]{http://astroparticles.es/}
\affiliation{\AddrAHEP}


\begin{abstract}
A realistic five-dimensional warped scenario with all \sm fields propagating in the bulk is proposed. Mass hierarchies would in principle be accounted for by judicious choices of the bulk mass parameters, while fermion mixing angles are restricted by a $\Delta(27)$ flavor symmetry broken on the branes by flavon fields.
The latter gives stringent predictions for the neutrino mixing parameters, and the Dirac CP violation phase, all described in terms of only two independent parameters at leading order. The scheme also gives an adequate CKM fit and should be testable within upcoming oscillation experiments.
\end{abstract}

\maketitle

\section{Introduction}
\label{sec:introduction}

The understanding of flavor constitutes one of the most stubborn open
challenges in particle physics~\cite{Agashe:2014kda}. Two aspects of
the problem are the understanding of fermion mass hierarchies as well
as mixing parameters.
Various types of flavor symmetries have been invoked in this
context~\cite{Babu:2002dz,Morisi:2012fg,Altarelli:2012bn,Fonseca:2014lfa,Altarelli:2010gt,Ishimori:2010au,King:2013eh,King:2014nza}. These
efforts have been partly motivated by the original success of the
tri-bimaximal mixing ansatz~\cite{Harrison:2002er}. The resulting
non-Abelian flavor symmetries are typically broken spontaneously down
to two different residual subgroups in the neutrino and the charged
lepton sectors, leading to zero reactor mixing parameter, $
\theta_{13} = 0$ .
However, the measurement of a non-zero value for the reactor
angle~\cite{An:2012eh,Abe:2011sj,Adamson:2013whj,Ahn:2012nd} implies
the need to revamp the original flavor symmetry-based approaches in
order to generate $ \theta_{13} \neq 0$~\cite{Morisi:2013qna} or else
look for alternative possibilities, such as bi-large neutrino
mixing~\cite{Boucenna:2012xb,Ding:2012wh,Roy:2012ib}.

The existence of warped extra-dimensions has been advocated by Randall
\& Sundrum~\cite{Randall:1999ee} as a way to address the
hierarchy problem, since the fundamental scale of gravity is
exponentially reduced from the Planck mass down to the TeV scale as a
result of having the Higgs sector localized near the boundary of the
extra dimensions.
Moreover, if \sm fermions are allowed to propagate in the bulk and
also become localized towards either brane, the scenario can also
address the flavor problem possibly acting in synergy with the flavor
group predictions. This is what we do in the present paper.

The idea of combining discrete flavor symmetries and extra dimensions
is quite attractive and has already been discussed in the literature
within the context of large extra
dimensions~\cite{Altarelli:2005yp,Altarelli:2008bg,Burrows:2009pi}, warped extra
dimensions~\cite{Csaki:2008qq,Chen:2009gy,Kadosh:2010rm,Kadosh:2011id,Kadosh:2013nra} and holographic composite Higgs
models~\cite{delAguila:2010vg,Hagedorn:2011pw,Hagedorn:2011un}. However, such models try to generate
tri-bimaximal neutrino mixing, which has been ruled out by the
measurement of the reactor angle
$\theta_{13}$~\cite{An:2012eh,Abe:2011sj,Adamson:2013whj,Ahn:2012nd}
and also global fits of neutrino oscillation
data~\cite{Forero:2014bxa}. One of us has constructed a warped extra
dimension model with $S_4$ flavor symmetry where democratic mixing is
produced at leading order and non-zero $\theta_{13}$ can arise from
subleading corrections~\cite{Ding:2013eca}. In this work, we shall
re-consider the issue of predicting flavor properties in particle
physics by combining the conventional predictive power inherent in the
use of non-Abelian flavor symmetries with the presence of warped
extra-dimensions.
We propose a warped five-dimensional scenario in which all matter
fields propagate in the bulk and neutrinos are treated as Dirac
particles. Our model can accommodate all the strengths of the \sm
Yukawa couplings and resulting fermion mass hierarchies by making
adequate choices of fermion bulk mass parameters, while the fermion
mixing parameters can be restricted by means of the assumed flavor
symmetry. We present a $\Delta(27)$ based flavor symmetry which nicely
describes the neutrino oscillation parameters in terms of just two
independent parameters, leading to interesting correlations involving
the neutrino mass hierarchy and the leptonic Dirac CP phase, not yet
reliably determined by current global oscillation
fit~\cite{Forero:2014bxa}. Our predictions include a neat leading
order relation between the solar and reactor mixing parameters which
should be tested at future oscillation experiments.

\section{Basic structure of the model}
\label{sec:structure}

In this section we present the basic setup of a warped
five-dimensional (5D) model for fermions, constructed under a
$\Delta(27)\otimes Z_4\otimes Z^{\prime}_4$ flavor symmetry.
The 5D field theory is defined on a slice of $\text{AdS}_5$, where the bulk
geometry is described by the metric
\begin{equation}
ds^2=e^{-2ky}\eta_{\mu\nu}dx^{\mu}dx^{\nu}-dy^2 \,,
\end{equation}
with $\eta_{\mu\nu}=\text{diag} (1,-1,-1,-1)$ and $k$ as the $\text{AdS}_5$
curvature scale. The fifth dimension $y$ is compactified on $S_1/Z_2$, and
two flat 3-branes of opposite tension are attached to the orbifold
fixed points, located at $y=0$ (UV brane) and $y=L$ (IR brane).

The electroweak symmetry of the model is promoted to $G_{\text{bulk}}=
SU(2)_L\otimes SU(2)_{R}\otimes U(1)_{B-L}$ in order to avoid
excessive contributions to the Peskin-Takeuchi $T$ parameter~\cite{Agashe:2003zs,Agashe:2006at}. The gauge group $G_{\text{bulk}}$
breaks down to the \sm electroweak (EW) group
$G_{\text{SM}}=SU(2)_{L}\otimes U(1)_{Y}$ on the UV brane by the
boundary conditions (BCs) of the gauge bosons. Furthermore, a bulk
Higgs field with $(SU(2)_L,SU(2)_{R})$ quantum numbers
\begin{equation}
H\sim (\mathbf{2},\mathbf{2})
\end{equation}
is responsible for the spontaneous symmetry breaking (SSB) of
$G_{\text{SM}}$.  The 5D Higgs field $H(x^\mu,y)$ can be decomposed into
Kaluza-Klein (KK) modes as
\begin{equation}
H(x^\mu,y)= H(x^\mu) \frac{f_H(y)}{\sqrt{L}} + \text{heavy KK Modes}\,.
\end{equation}
For an adequate choice of BCs, its zero mode
profile $f_H(y)$ can be written as~\cite{Cacciapaglia:2006mz}
\begin{equation}
  f_H(y)=\sqrt{\frac{2 k L (1-\beta )}{1-e^{-2(1-\beta )k L}} } e^{kL}e^{(2-\beta)k(y-L)}\,,
\end{equation}
where we have introduced the Higgs localization parameter
$\beta=\sqrt{4+m^2_H/k^2}$ in terms of the Higgs field bulk mass
parameter $m_H$. In the present work, we assume that the vacuum
expectation value (VEV) of the Higgs zero mode is of the form
\begin{equation}
\langle H(x^\mu)\rangle = \frac{v_H}{\sqrt{2}}
\left(
\begin{array}{cc}
1 & 0  \\
0 & 1
\end{array}
\right)
\,,
\end{equation}
and it is peaked toward the IR brane,
allowing for a TeV scale EW SSB and inducing the $G_{\text{bulk}}$ breakdown to
$SU(2)_{D}\otimes U(1)_{B-L}$ on that brane.

Three families of fermion fields are required to describe each
generation (labeled by $i=1,2,3$) of quarks and leptons.  All fermion
fields propagate into the bulk and transform under the minimal
representation of the gauge group $SU(2)_L\otimes SU(2)_{R}$
\cite{Agashe:2003zs,Agashe:2006at}.
In the lepton sector the three multiplets of the model are given as
\begin{equation}
\Psi_{\ell_i}=\left(\begin{array}{c}
\nu_{ i}^{[++]}\\
e_{i}^{[++]}
\end{array}\right)\sim (\mathbf{2},\mathbf{1})\,,\qquad
\Psi_{e_i}=\left(\begin{array}{c}
\nu_{i}^{[+-]}\\
e_{ i}^{[--]}
\end{array}\right)\sim(\mathbf{1},\mathbf{2})\,,\qquad
\Psi_{\nu_i}=\left(\begin{array}{c}
\nu_{i}^{[--]}\\
e_{i}^{[+-]}
\end{array}\right)\sim(\mathbf{1},\mathbf{2})\,,
\end{equation}
while for the quark sector we have
\begin{equation}
\Psi_{Q_i}=\left(\begin{array}{c}
u_{i}^{[++]}\\
d_{i}^{[++]}
\end{array}\right)\sim (\mathbf{2},\mathbf{1})\,,\qquad
\Psi_{d_i}=\left(\begin{array}{c}
u_{i}^{[+-]}\\
d_{i}^{[--]}
\end{array}\right)\sim(\mathbf{1},\mathbf{2})\,,\qquad
\Psi_{u_i}=\left(\begin{array}{c}
u_{i}^{[--]}\\
d_{i}^{[+-]}
\end{array}\right)\sim(\mathbf{1},\mathbf{2})\,.
\end{equation}
Notice that we have a separate $SU(2)_{R}$ doublet for every right
handed fermion. In the above equations, fields with different sign
assignments must be understood as independent. The bracketed signs
indicate Neumann ($+$) or Dirichlet ($-$) BCs for the left-handed
component of the corresponding field, on both UV and IR branes. The
right-handed part of the field satisfies opposite BCs. Only fields
with $[++]$ BCs have left-handed zero modes, whereas right-handed zero
modes exist solely for fields with $[--]$ BCs. The KK decomposition
for such fields has the form
\begin{eqnarray}
&&\psi^{[++]}(x^\mu,y)  =   \frac{e^{2ky}}{\sqrt{L}}  \Big\{ \psi_L(x^\mu) f_L^{(0)}(y,c_{L}) +\text{heavy KK modes}  \Big\}\,,
\\
&&\psi^{[--]}(x^\mu,y) =  \frac{e^{2ky}}{\sqrt{L}}  \Big\{\psi_{R}(x^\mu) f_R^{(0)}(y,c_{R}) + \text{heavy KK modes}  \Big\}\,,
\nn
\label{FermionKK}
\end{eqnarray}
with $\psi=\nu_i,e_i,u_i,d_i$, and zero mode profiles  \cite{Gherghetta:2000qt,Grossman:1999ra,Huber:2001ug}
\begin{equation}
f_L^{(0)}(y,c_{L})=\sqrt{\frac{(1-2 c_{L})kL}{e^{(1-2c_{L})kL}-1}} e^{-c_{L}ky}\,,
\hspace{1cm}
f_R^{(0)}(y,c_R)=\sqrt{\frac{(1+2c_R)kL}{e^{(1+2c_R)kL}-1}} e^{ c_R ky}\,,
\end{equation}
where $c_{L}$ and $c_{R}$ are the bulk mass parameters of the 5D
fermion fields in units of the $\text{AdS}_5$ curvature $k$. Thus, the
low energy spectrum contains left-handed doublets
$\ell_{iL}=(\nu_{iL},e_{iL})$, $Q_{iL}=(u_{iL},d_{iL})$, alongside
right-handed singlets $\nu_{iR},e_{iR},u_{iR},d_{iR}$. In the
following, we identify all \sm fields with this set of zero modes ({\it i.e.} the so called zero
mode approximation, ZMA). For future convenience, we denote the flavor
components of charged leptons and quarks as $e_{1,2,3}=e,\mu,\tau$;
$Q_{1,2,3}=U,C,T$; $u_{1,2,3}=u,c,t$; $d_{1,2,3}=d,s,b$.

In the present work, we choose the flavor symmetry to be $\Delta(27)$, augmented by the auxiliary symmetry $Z_4\otimes Z^{\prime}_4$. The group $\Delta(27)$ was originally proposed to explain the fermion masses and flavor mixing in Refs.~\cite{deMedeirosVarzielas:2006fc,Ma:2006ip}, and has been used for Dirac neutrinos in \cite{Aranda:2013gga} by one of us. Here we study its implementation in a warped extra dimensional theory. The flavor symmetry $\Delta(27)\otimes Z_4\otimes Z_4^\prime$ is broken by brane localized flavons, transforming as singlets under $G_{\text{bulk}}$. We introduce a set of flavons $\xi$, $\sigma_1$, $\sigma_2$ localized on the IR brane, and a flavon $\varphi$ localized on the UV brane. Both $\xi$ and $\varphi$ are assigned to the three dimensional representation $\mathbf{3}$ of $\Delta(27)$, while $\sigma_1$ and $\sigma_2$ transform as inequivalent one dimensional representations $\mathbf{1}_{0,1}$ and $\mathbf{1}_{0,0}$ respectively. A summary of the $\Delta(27)$ group properties and its representations can be found in Appendix \ref{sec:App_A}. There are two different scenarios for the model, determined by the two possible VEV alignments for $\xi$, namely:
\begin{equation}
\begin{split}
\langle \xi \rangle&=(0,1,0)v_\xi,\qquad \text{Case I},\\
\langle \xi \rangle&=(1,\omega,1)v_\xi,\qquad \text{Case II},
\end{split}
\label{Cases}
\end{equation}
with $\omega=e^{2\pi i/3}$. As indicated above, we will denote the
models described by each alignment as cases I and II,
respectively. Note that the case II vacuum pattern frequently appears in the context of geometrical
CP violation~\cite{Branco:1983tn,Bhattacharyya:2012pi}. The VEVs for the remaining flavon fields are
\begin{equation}
\langle \varphi \rangle=(1,1,1)v_\varphi\,,
\hspace{0.6cm}
\langle \sigma_1 \rangle=v_{\sigma_1}\,,
\hspace{0.6cm}
\langle \sigma_2 \rangle=v_{\sigma_2}\,.
\label{FlavonVEV}
\end{equation}
Further details regarding this vacuum configuration are offered in
Appendix \ref{sec:vacuum_alignment}.

\section{Lepton sector}
\label{lepton}

Once the basic framework has been laid out, we are in position to
discuss the structure of the lepton sector and its phenomenological
implications. As we will show below, charged lepton as well as Dirac
neutrino masses are generated at leading order (LO), and non-zero
values for the ``reactor angle'' $\theta_{13}$ arise naturally. The
model is predictive, in the sense that the three mixing angles and the
Dirac CP phase will ultimately be determined in terms of only two
parameters.

\subsection{Lepton masses and mixing}
\label{sec:Lmm}

\begin{table} [ht]
\centering
\begin{tabular}{|c||c|c|c|c|c|c|c|c||c||c|c|c|}
\hline \hline
Field      &  $\Psi_\ell$ &  $\Psi_e$     &   $\Psi_\mu$    &      $\Psi_\tau$      &
           $\Psi_{\nu_1}$  &       $\Psi_{\nu_2}$        &      $\Psi_{\nu_3}$   &
           $H$  &  $\varphi$  &          $\xi$              &        $\sigma_1$     &    $\sigma_2$
\\ \hline
$\Delta(27)$  &  $\mathbf{3}$  &   $\mathbf{1}_{0,0}$ &  $\mathbf{1}_{1,0}$ &  $\mathbf{1}_{2,0}$ &
            $\mathbf{1}_{0,0}$ &   $\mathbf{1}_{0,0}$ &  $\mathbf{1}_{0,0}$ &
            $\mathbf{1}_{0,0}$ &      $\mathbf{3}$    &     $\mathbf{3}$    &   $\mathbf{1}_{0,1}$ &  $\mathbf{1}_{0,0}$
\\ \hline
$Z_4$    &  $1$    &  $1$  &    $1$  &   $1$  &
         $-1$  &  $i$  &     $-1$   &
             1     &      $1$  &  $-1$  &     1       &     $i$
\\\hline
$Z_4^\prime$    & 1      &  $i$  &    $i$  &  $i$  &
           $-1$   &  $-1$ &    $-1$ &
           1      &  $-i$ &    $1$  &  $-1$  &   $-1$
\\ \hline
\hline
\end{tabular}
\caption{\label{tl} Particle content and transformation
  properties of the lepton and scalar sectors under the flavor symmetry $\Delta(27) \otimes  Z_4\otimes Z_4^\prime$.}
\end{table}

The transformation properties of leptons and scalars under the family
symmetry $\Delta(27) \otimes Z_4\otimes Z_4^\prime$ are given in Table~\ref{tl}. Note that the Higgs field is inert under the flavor
symmetry. Since the three left-handed lepton doublets are unified into
a faithful triplet $\mathbf{3}$ of $\Delta(27)$, they will share one
common bulk mass parameter $c_{\ell}$. On the other hand, both
right-handed charged leptons and right-handed neutrinos are assigned
to singlet representations of $\Delta(27)$. Therefore, there are six
different bulk mass parameters $c_{e_i}$ and $c_{\nu_{i}}$ ($i=1, 2,
3$) for these fields.
From the particle transformation properties we can write the most
general lepton Yukawa interactions that are both gauge and flavor
invariant at LO:
\begin{eqnarray}
{\cal L}^{l}_Y &=&
\frac{\sqrt{G}}{ \Lambda^{\frac{5}{2}} }
\Bigg\{
y_e    \big(\varphi \overline{\Psi}_\ell\big)_{\mathbf{1}_{0,0}}   H \Psi_e    +
y_\mu  \big(\varphi \overline{\Psi}_\ell\big)_{\mathbf{1}_{2,0}}   H \Psi_\mu  +
y_\tau \big(\varphi \overline{\Psi}_\ell\big)_{\mathbf{1}_{1,0}}   H \Psi_\tau
\Bigg\}\delta(y)
\nonumber\\
&&+
 \frac{\sqrt{G}}{ (\Lambda^\prime)^{\frac{7}{2}} }  \Bigg\{
y_{11}  \big(\xi \sigma_1   \overline{\Psi}_\ell\big)_{\mathbf{1}_{0,0}}  \widetilde{H} \Psi_{\nu_1}  +
y_{31}  \big(\xi \sigma_1^* \overline{\Psi}_\ell\big)_{\mathbf{1}_{0,0}}  \widetilde{H} \Psi_{\nu_1}  +
y_{22}     \big(\xi \sigma_2   \overline{\Psi}_\ell\big)_{\mathbf{1}_{0,0}}  \widetilde{H} \Psi_{\nu_2}
\nonumber\\&&\hspace{1cm}+
y_{13}  \big(\xi \sigma_1   \overline{\Psi}_\ell\big)_{\mathbf{1}_{0,0}}  \widetilde{H} \Psi_{\nu_3}  +
y_{33}  \big(\xi \sigma_1^* \overline{\Psi}_\ell\big)_{\mathbf{1}_{0,0}}  \widetilde{H} \Psi_{\nu_3}
\Bigg\}\delta(y-L)
+
\text{h.c.}
\label{eq:wl_LO}
\end{eqnarray}
with $\widetilde{H}\equiv\tau_2 H^{*}\tau_2$, and $\tau_{i}$ as the
Pauli matrices.  After electroweak and flavor spontaneous symmetry
breaking, all leptons develop masses dictated by the above Yukawa
interactions. The generated masses are modulated by the overlap of the
relevant zero mode fermion profiles, the VEV profile of the Higgs, and
the flavon VEVs given in Eqs.~(\ref{Cases}, \ref{FlavonVEV}).

From Eq.~\eqref{eq:wl_LO}, The mass matrix $m_l$ for charged leptons is
\begin{equation}
m_l = \frac{1}{(L\,\Lambda)^\frac{3}{2}} \frac{v_\varphi}{\Lambda}
\frac{v}{\sqrt{2}} \sqrt{3} \, U_{l} \left(
\begin{array}{ccc}
\widetilde{y}_e  &  0                  &    0                 \\
0                &  \widetilde{y}_\mu  &    0                 \\
0                &  0                  &   \widetilde{y}_\tau
\end{array}
\right)\,,
\label{mch}
\end{equation}
where $U_{l}$ stands for the so-called magic matrix
\begin{equation}
U_{l}
=
\frac{1}{\sqrt{3}}
\left(\begin{array}{ccc}
1&1&1\\
1&\omega&\omega^2\\
1&\omega^2&\omega
\end{array}\right)\,,
\end{equation}
 and $\widetilde{y}_{e,\mu,\tau}$ are modified Yukawa couplings defined as
\begin{eqnarray}
\widetilde{y}_{e,\mu,\tau}    &=&  y_{e,\mu,\tau} F(0,c_\ell,c_{e_i})\,,
\end{eqnarray}
in terms of the overlapping function
\begin{eqnarray}
F(y,c_{L},c_R) &\equiv & f^{(0)}_L(y,c_{L}) \, f^{(0)}_R(y,c_R) f_H(y)
\nn\\&=&
\sqrt{\frac{2\left(1-\beta_H\right)\left(1-2c_{L}\right)\left(1+2c_R\right)k^3L^3}{\left[1-e^{-2(1-\beta_H)kL}\right]\left[e^{(1-2c_{L})kL}-1\right]\left[e^{(1+2c_R)kL}-1\right]}}\;e^{-(1-\beta_H)kL}e^{(2-\beta_H-c_{L}+c_R)ky}\,.
\end{eqnarray}
Given that $U_l^{\dagger}U_l=1$, the diagonalization of the charged
lepton mass matrix is straightforward, leading to charged lepton
masses of the form
\begin{equation}
m_{e,\mu,\tau}    = \frac{\sqrt{3}\, \widetilde{y}_{e,\mu,\tau}   }{ (L\,\Lambda)^\frac{3}{2} }  \frac{v_\varphi}{\Lambda}\frac{v}{\sqrt{2}}\,.
\end{equation}

Analogously, taking into account the two distinct VEV alignments for
the flavon triplet $\xi$ in Eq.~(\ref{Cases}), the neutrino mass matrix
for each respective case can be written as
\begin{equation}
m^{\text{I}}_\nu=
\frac{1}{ (L\,\Lambda^\prime)^\frac{3}{2} }
\frac{v_\xi}{\Lambda^\prime}
\frac{v}{\sqrt{2}}
\left(
\begin{array}{ccc}
\widetilde{y}_{11} \frac{v_{\sigma_1}}{\Lambda^\prime}  &  0  &  \widetilde{y}_{13} \frac{v_{\sigma_1}}{\Lambda^\prime}  \\
0                                                       &  \widetilde{y}_{22} \frac{v_{\sigma_2}}{\Lambda^\prime}  &  0  \\
\widetilde{y}_{31} \frac{v_{\sigma_1}^\ast}{\Lambda^\prime}  &  0  &  \widetilde{y}_{33} \frac{v_{\sigma_1}^\ast}{\Lambda^\prime}
\end{array}
\right)\,,
\label{mnuI}
\end{equation}
\begin{equation}
m^{\text{II}}_\nu=
\frac{1}{ (L\,\Lambda^\prime)^\frac{3}{2} }
\frac{v_\xi}{\Lambda^\prime}
\frac{v}{\sqrt{2}}
\sqrt{3} V_0
\left(
\begin{array}{ccc}
\widetilde{y}_{11} \frac{v_{\sigma_1}}{\Lambda^\prime}  &  0  &  \widetilde{y}_{13} \frac{v_{\sigma_1}}{\Lambda^\prime}  \\
0                                                       &  \widetilde{y}_{22} \frac{v_{\sigma_2}}{\Lambda^\prime}  &  0  \\
\widetilde{y}_{31} \frac{v_{\sigma_1}^\ast}{\Lambda^\prime}  &  0  &  \widetilde{y}_{33} \frac{v_{\sigma_1}^\ast}{\Lambda^\prime}
\end{array}
\right)\,,
\label{mnuII}
\end{equation}
with
\begin{eqnarray}
\widetilde{y}_{ij} &=& y_{ij}  F(L,c_\ell,c_{\nu_j}) \,,
\end{eqnarray}
and
\begin{eqnarray}
V_0&\equiv&
\frac{1}{\sqrt{3}}
\left(
\begin{array}{ccc}
\omega     &      1        &       1         \\
1          &    \omega     &       1         \\
1          &      1        &     \omega      \\
\end{array}
\right)\,.
\end{eqnarray}
Thus, the diagonalizing matrix for the neutrino sector can be parameterized as
\begin{equation}
U^{\text{I}}_\nu=
\left(\begin{array}{ccc}
\cos{\theta_\nu}                     &    0    &\sin{\theta_\nu} e^{i\varphi_\nu}\\
0                                    &    1    &                0                 \\
-\sin{\theta_\nu} e^{-i\varphi_\nu}  &    0    &\cos{\theta_\nu}
\end{array}\right)\,,
\label{thetanuI}
\end{equation}
\begin{equation}
U^{\text{II}}_\nu=
V_0
\left(\begin{array}{ccc}
\cos{\theta_\nu}                     &    0    &\sin{\theta_\nu} e^{i\varphi_\nu}\\
0                                    &    1    &                0                 \\
-\sin{\theta_\nu} e^{-i\varphi_\nu}  &    0    &\cos{\theta_\nu}
\end{array}\right)\,.
\label{thetanuII}
\end{equation}
In terms of the auxiliary functions
\begin{equation}
\label{XY}
X^{\pm}_\nu=|\widetilde{y}_{31}|^2+|\widetilde{y}_{33}|^2\pm|\widetilde{y}_{11}|^2\pm|\widetilde{y}_{13}|^2\,,\qquad
Y_\nu= \widetilde{y}_{11} \widetilde{y}_{33}-\widetilde{y}_{13}^\ast \widetilde{y}_{31}^\ast \,,\qquad Z_\nu= \widetilde{y}_{11} \widetilde{y}_{31}^\ast +\widetilde{y}_{13} \widetilde{y}_{33}^\ast\,,
\end{equation}
the relevant parameters of the model, $\theta_\nu$ and $\varphi_\nu$, are given by
\begin{equation}
\label{par1}
\tan{2\theta_\nu}=2|Z_\nu|/X^{-}_\nu,\qquad
\varphi_\nu=\arg \left(v_{\sigma_1}^2 Z_\nu \right)\,,
\end{equation}
and the neutrino mass eigenvalues for both NH and IH are determined as
\begin{itemize}
\item Case I
\begin{eqnarray}
\text{NH:}\hspace{0.4cm}
m_1=
\frac{\widetilde{v}_1}{\sqrt{2}}
M^-\big(X^{+}_\nu,Y_\nu\big)\,,
\hspace{0.2cm}
m_2=\widetilde{v}_2\left|\widetilde{y}_{22}\right|\,,
\hspace{0.2cm}
m_3=
\frac{\widetilde{v}_1}{\sqrt{2}}M^+\big(X^{+}_\nu,Y_\nu\big),\quad\mathrm{for}~ X^{-}_\nu\cos2\theta_\nu>0\,,  \\
\text{IH:}\hspace{0.4cm}
m_1=
\frac{\widetilde{v}_1}{\sqrt{2}}
M^+\big(X^{+}_\nu,Y_\nu\big)\,,
\hspace{0.2cm}
m_2=\widetilde{v}_2\left|\widetilde{y}_{22}\right|\,,
\hspace{0.2cm}
m_3=
\frac{\widetilde{v}_1}{\sqrt{2}}M^-\big(X^{+}_\nu,Y_\nu\big),\quad \mathrm{for}~X^{-}_\nu\cos2\theta_\nu<0\,,
\label{NHspectrum}
\end{eqnarray}
\item Case II
\begin{eqnarray}
\text{NH:}\hspace{0.4cm}
m_1=
\sqrt{\frac{3}{2}}\widetilde{v}_1
M^-\big(X^{+}_\nu,Y_\nu\big)\,,
\hspace{0.2cm}
m_2=\sqrt{3}\widetilde{v}_2\left|\widetilde{y}_{22}\right|\,,
\hspace{0.2cm}
m_3=
\sqrt{\frac{3}{2}}\widetilde{v}_1 M^+\big(X^{+}_\nu,Y_\nu\big),\quad\mathrm{for}~ X^{-}_\nu\cos2\theta_\nu>0\,,  \\
\text{IH:}\hspace{0.4cm}
m_1=
\sqrt{\frac{3}{2}}\widetilde{v}_1
M^+\big(X^{+}_\nu,Y_\nu\big)\,,
\hspace{0.2cm}
m_2=\sqrt{3}\widetilde{v}_2\left|\widetilde{y}_{22}\right|\,,
\hspace{0.2cm}
m_3=
\sqrt{\frac{3}{2}}\widetilde{v}_1 M^-\big(X^{+}_\nu,Y_\nu\big),\quad \mathrm{for}~X^{-}_\nu\cos2\theta_\nu<0\,,
\label{IHspectrum}
\end{eqnarray}
\end{itemize}
where we have defined
\begin{equation}
M^{\pm}(x,y)=\sqrt{x\pm\sqrt{x^2-4 |y|^2}}\,,
\label{Mpm}
\end{equation}
and
\begin{equation}
\label{eq:vtilde_def}\widetilde{v}_\alpha=\left|\frac{1}{(L\,\Lambda^\prime)^\frac{3}{2}}
\frac{v_\xi}{\Lambda^\prime}
\frac{v_{\sigma_{\alpha}}}{\Lambda^\prime}
\frac{v}{\sqrt{2}}\right|\,,\qquad \alpha=1,2\,.
\end{equation}

Without loss of generality, the angle $\theta_\nu$ is restricted to the interval $[0,\pi]$. Notice that $X^{-}_{\nu}\cos2\theta_\nu=2|Z_{\nu}|\cos^22\theta_{\nu}/\sin2\theta_{\nu}$. As a result, for non-vanishing values of $Z_\nu$, the neutrino mass spectrum displays Normal Hierarchy (NH) provided $0<\theta_\nu<\pi/2$, whereas Inverted Hierarchy (IH) is realized for $\pi/2<\theta_\nu<\pi$. The angle $\varphi_\nu$, on the other hand, can take any value in the interval $[0,2\pi]$.

At leading order, the lepton mixing matrix $U_{\text{PMNS}}=U_{l}^\dagger
U_{\nu}$ becomes
\begin{equation}
  U^{\text{I}}_{\text{PMNS}}= \frac{1}{\sqrt{3}} \left(
\begin{array}{ccc}
\cos\theta_{\nu}-e^{-i\varphi_{\nu}}\sin\theta_{\nu}
&
1
&
\cos\theta_{\nu}+e^{i\varphi_{\nu}}\sin\theta_{\nu}
\\
\cos\theta_{\nu}-\omega e^{-i\varphi_{\nu}}\sin\theta_{\nu}
&
\omega^2
&
\omega \cos\theta_{\nu}+e^{i\varphi_{\nu}}\sin\theta_{\nu}
\\
\cos\theta_{\nu}-\omega^2 e^{-i\varphi_{\nu}}\sin\theta_{\nu}
&
\omega
&
\omega^2\cos\theta_{\nu}+ e^{i\varphi_{\nu}}\sin\theta_{\nu}
\\
\end{array}
\right)\,,
\end{equation}
\begin{equation}
U^{\text{II}}_{\text{PMNS}}
=
\frac{-i \omega}{\sqrt{3}}
\left(
\begin{array}{ccc}
\cos \theta _{\nu }-e^{-i \varphi _{\nu }} \sin \theta _{\nu }                     &
1                                                                                 &
\cos \theta _{\nu }+e^{i \varphi_{\nu }} \sin \theta _{\nu }
\\
\omega \cos \theta _{\nu }- \omega^2 e^{-i \varphi _{\nu }}  \sin \theta _{\nu }   &
1                                                                                 &
\omega^2  \cos \theta _{\nu }+ \omega e^{i \varphi _{\nu }} \sin \theta _{\nu }
\\
\omega  \cos \theta _{\nu }-e^{-i \varphi _{\nu }} \sin \theta _{\nu }             &
\omega ^2                                                                         &
\cos \theta_{\nu}+ \omega e^{i \varphi _{\nu }} \sin \theta _{\nu } \\
\end{array}
\right)\,.
\end{equation}
In both cases, the solar, atmospheric and reactor angles can be
written in terms of $\theta_\nu$ and $\varphi_\nu$ as
\begin{eqnarray}
\sin^2\theta_{12}&=&\frac{1}{2-\sin2\theta_\nu\cos\varphi_\nu} \,,
\nn\\
\sin^2\theta_{23}&=&\frac{1-\sin2\theta_\nu\sin(\pi/6-\varphi_\nu)}{2-\sin2\theta_\nu\cos\varphi_\nu}
\,,
\nn\\
\sin^2\theta_{13}&=&\frac{1}{3}\left( 1+\sin2\theta_\nu\cos\varphi_\nu
\right) \,.
\label{angles}
\end{eqnarray}
A convenient description for the CP violating phase in this sector is
the Jarlskog invariant $J_{\text{CP}}=\mathbf{Im}[U^{*}_{e
  1}U^{*}_{\mu 3}U_{\mu 1}U_{e 3}]$~\cite{Jarlskog:1985ht}, which in
this parameterization takes the compact form
\begin{eqnarray}
J_{\text{CP}}&=&-\frac{1}{6\sqrt{3}}\cos 2\theta_\nu \,.
\label{JCP}
\end{eqnarray}
It is worthy of attention the independence of $J_{\text{CP}}$ upon
$\varphi_{\nu}$, and the simple predicted relation between the solar
and reactor angles $\theta_{12}$ and $\theta_{13}$:
\begin{eqnarray}
\sin^2\theta_{12}\cos^2\theta_{13}&=&\frac{1}{3}\,.
\end{eqnarray}

\subsection{Phenomenological implications}
\label{sec:pheno}

As shown above, only two parameters are required to generate the three
angles and the Dirac CP violating phase characterizing the lepton
mixing matrix, making this model highly predictive. In the remaining
part of this section we explore in detail the predictions for the
lepton mixing parameters and the neutrino mass spectrum.
\begin{figure}[t!]
\centering
\includegraphics[scale=0.8]{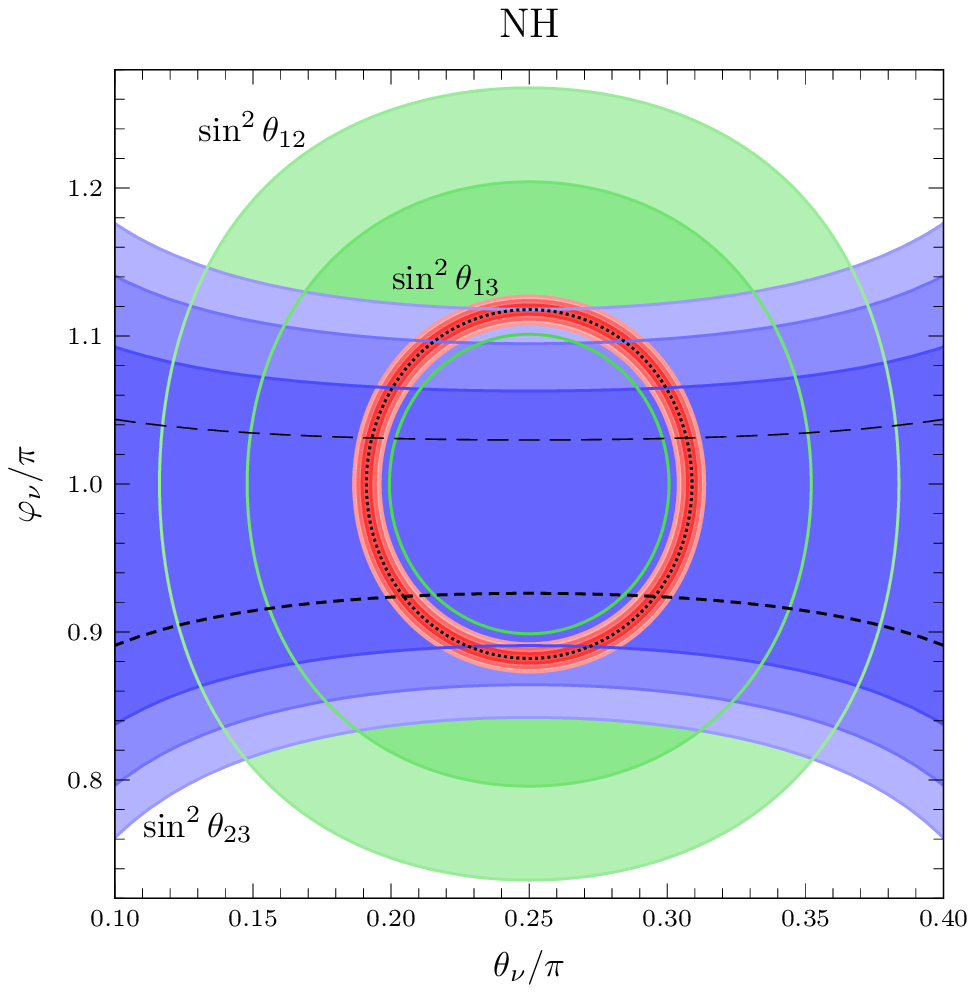}
\includegraphics[scale=0.8]{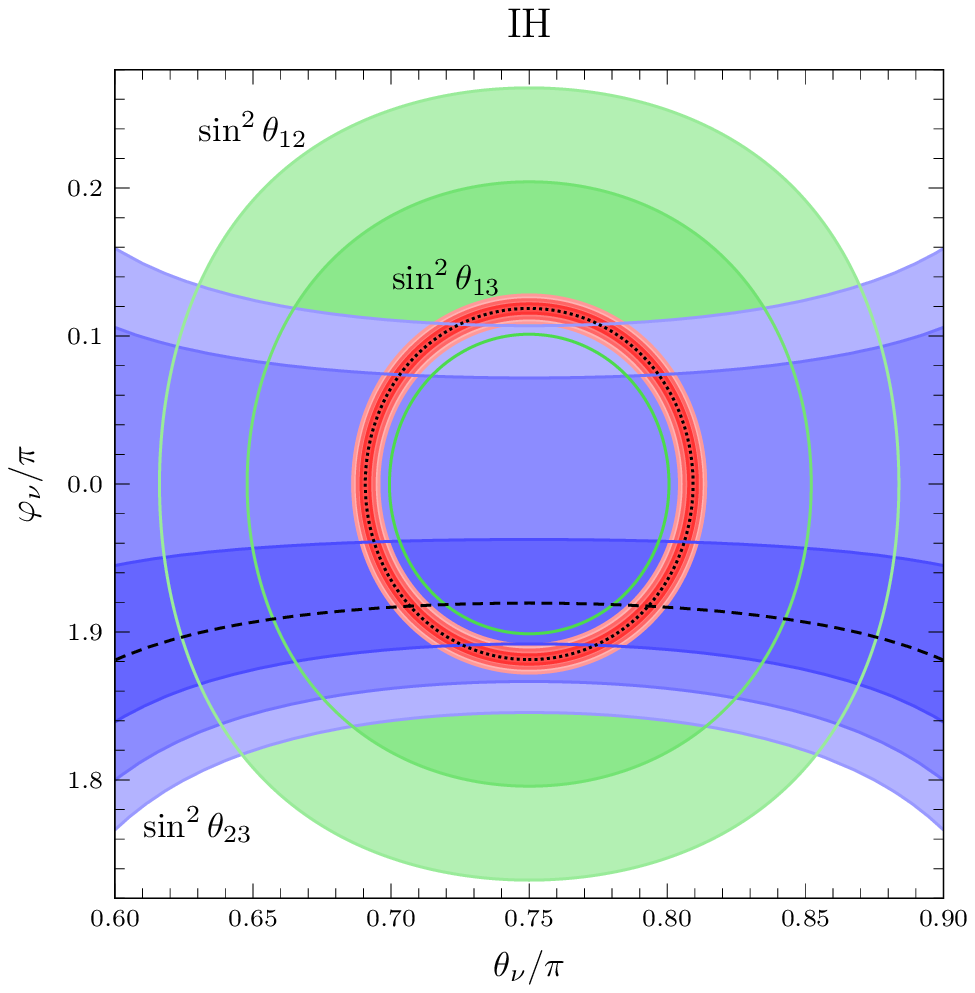}
\caption{\small{1$\sigma$, 2$\sigma$ and 3$\sigma$ ranges of $\sin^2\theta_{12}$ (green), $\sin^2\theta_{23}$ (blue) and $\sin^2\theta_{13}$ (red) for normal (left panel) and inverted
(right panel) neutrino mass hierarchies. Best-fit contours for
$\sin^2\theta_{13}$ ($\sin^2\theta_{23}$) are indicated by dotted
(short--dashed) lines. The long--dashed contour in the left panel
represents the local minimum in the first octant of $\theta_{23}$.}}
\label{fig:NH_IH}
\end{figure}

In Figure~\ref{fig:NH_IH}, the $\theta_\nu$ -- $\varphi_\nu$ parameter
region compatible with experimental data is delimited using the global
fit of neutrino oscillations given in~\cite{Forero:2014bxa} for each
mass ordering, shown as the left and right hand panel.  The model can
reproduce successfully the best-fit values for the atmospheric and
reactor angles, reaching simultaneously the 2$\sigma$ region for the
solar angle. The intersecting points of the ``central'' or best fit
curve in the $\sin^2\theta_{13}$ contour and the corresponding ones in
the $\sin^2\theta_{23}$ contour are located at
\begin{equation}\label{Theta-Nu}
\begin{array}{ccccc}
\text{$\mathrm{NH}_{1}$}: &\qquad & \theta_\nu/\pi=0.204 (0.296)\,, &\qquad & \varphi_\nu/\pi=0.924\,,\\
\text{$\mathrm{NH}_{2}$}: &\qquad & \theta_\nu/\pi=0.193 (0.307)\,, &\qquad & \varphi_\nu/\pi=1.031\,,\\
\text{IH}: &\qquad &\theta_\nu/\pi=0.707 (0.793)\,, &\qquad &\varphi_\nu/\pi=1.917\,,
\end{array}
\end{equation}
where $\mathrm{NH}_{1}$ denotes the best-fit contour of
$\sin^2\theta_{23}$, and $\mathrm{NH}_{2}$ corresponds to its local
minimum in the first octant. Notice that the numbers in
parenthesis denote the intersection values within the range
$\theta_{\nu} \in [\pi/4, \pi/2] \cup  [3\pi/4, \pi] $.

Once we have determined $\theta_\nu$ and $\varphi_\nu$ from the
central values of the atmospheric and reactor oscillation global fits,
the predictions for the solar angle and the Jarlskog invariant can be
straightforwardly obtained using Eqs.~(\ref{angles}, \ref{JCP}). For
completeness, in Table~\ref{pred} we present the full set of mixing
parameters derived from the points defined in Eq.~(\ref{Theta-Nu}).

\begin{table} [ht]
\centering
\begin{tabular}{|c||c|c|c|c|}
\hline \hline
     &  $\mathrm{NH}_{1}$  &  $\mathrm{NH}_{2}$  &  IH
\\ \hline
$\sin^2\theta_{23}/10^{-1}$     &  $5.67$ & $4.73$ &  $5.73$
\\ \hline
$\sin^2\theta_{13}/10^{-2}$     &  $2.26$ & $2.26$ &  $2.29$
\\ \hline \hline
$\sin^2\theta_{12}/10^{-1}$     &  $3.41$ & $3.41$ & $3.41$
\\ \hline
$J_{\text{CP}}/10^{-2}$     & $-(+)2.71$ & $-(+)3.37$ & $+(-)2.57$
\\ \hline
 \hline
\end{tabular}
\caption{\label{pred} Central predictions for $\sin^2\theta_{12}$ and $J_{\text{CP}}$ obtained from the central values of the atmospheric and reactor angles reported in Ref.~\cite{Forero:2014bxa}. The sign of $J_{CP}$ in the parentheses corresponds to the bracketed prediction for $\theta_{\nu}$ in  Eq.~\eqref{Theta-Nu}.}
\end{table}

Remarkably, the central prediction for $\sin^2\theta_{12}$ falls very
close to its $1\sigma$ boundary. In addition, notice that the
$1\sigma$ range of $J_{\text{CP}}$ is entirely contained in the region
$\theta_\nu\in[0, \pi/4]\cup[3\pi/4, \pi]$.

We conclude this section bringing forth a consistent realization of
lepton masses and mixing angles. In the numerical analysis, we assume
that the fundamental 5D scale is $k\simeq\Lambda\simeq M_{\text{Pl}}$, with
$M_{\text{Pl}}\simeq2.44\times10^{18}$ GeV as the reduced Planck mass. We
also set the scale $\Lambda^\prime\simeq k'=ke^{-kL}\simeq 1.5$ TeV in
order to account for the hierarchy between the Planck and the
electroweak scales, allowing for the lowest KK gauge boson resonances
(with masses $m_{KK}=3\sim4$ TeV) to be within reach of the LHC
experiments. The Higgs VEV is identified with its \sm value
$v\simeq246$ GeV, and the ratios $v_\varphi/\Lambda$,
$v_\xi/\Lambda^\prime$, $v_{\sigma_1}/\Lambda^\prime$,
$v_{\sigma_2}/\Lambda^\prime$, are all fixed to 0.1 (thus considering
real-valued flavon VEVs). The Higgs localization parameter $\beta$,
common to all mass matrix elements, is chosen as $0.95$ in the
following discussion.

As an as illustrative example, we can choose $c_{\ell}=1.85$, $c_{e}=-0.27$, $c_{\mu}=-0.44$, $c_{\tau}=-0.71$, $|y_{e}|=0.861$, $|y_{\mu}|=0.898$, $|y_{\tau}|=0.994$ to generate the charged lepton
masses $m_e=0.511\,\mathrm{MeV}$, $m_{\mu}=105.7\,\mathrm{MeV}$,
$m_{\tau}=1.777\,\mathrm{GeV}$. For the neutrino sector, benchmark
points (BPs) in parameter space are given in Table~\ref{BP}. There,
the four BPs are labeled according to their hierarchy scheme and case
as NH-I, NH-II, IH-I, IH-II. One sees that, indeed, the large
disparity between charged lepton masses is reproduced for Yukawa
couplings of the same order of magnitude.
\begin{table} [ht]
\centering
\begin{tabular}{|c||c|c|c|c|}
\hline \hline
  & NH-I & NH-II     & IH-I & IH-II
\\ \hline
$c_{\nu_1}$ & $-1.40$ & $-1.41$     & $-1.39$ & $-1.40$
\\ \hline
$c_{\nu_2}$ & $-1.38$ & $-1.40$     & $-1.33$ & $-1.35$
\\ \hline
$c_{\nu_3}$ & $-1.34$ & $-1.36$     & $-1.34$ & $-1.36$
\\ \hline \hline
$y_{11}$ & $-1.000 - 0.307 i$ & $0.282 + 1.166 i$     & $0.752 + 0.096 i$  & $-0.674 + 0.520 i$
\\ \hline
$y_{13}$ & $-0.451 + 0.631 i$ & $0.031 - 0.880 i$    & $0.919 - 0.432 i$  & $1.026 - 0.542 i$
\\ \hline
$y_{22}$ & $0.860 + 0.353 i$ & $0.097 - 1.088 i$      &$-0.905 - 0.194 i$  & $0.974 + 0.431 i$
\\ \hline
$y_{31}$ & $0.667 + 0.397 i$ & $0.001 - 0.881 i$      & $0.941 + 0.383 i$   & $-1.070 + 0.450 i$
\\ \hline
$y_{33}$ & $0.792 - 0.683 i$ & $-0.324 + 1.154 i$     & $0.746 - 0.136 i$   & $0.829 - 0.191 i$
\\ \hline
 \hline
\end{tabular}
\caption{\label{BP} Benchmark points for the neutrino sector, featuring
both NH and IH in Cases I and II. }
\end{table}

The neutrino masses, splittings and mixing angles associated to each
BP are displayed in Table~\ref{NuMass}. All the obtained neutrino
oscillation parameters are consistent with the global fit in
Ref.~\cite{Forero:2014bxa}. In particular, the reproduced atmospheric
and reactor angles lie comfortably in their respective $1\sigma$
region, whereas the solar angle values are contained in the $2\sigma$
range, very close to the $1\sigma$ boundary.

\begin{table} [ht]
\centering
\begin{tabular}{|c||c|c|c|c|}
\hline \hline
     &  NH-I &  NH-II  &  IH-I &  IH-II
\\ \hline \hline
$m_1\,[\mathrm{eV}]$     & $ 1.80\times 10^{-3}$  & $ 2.59\times 10^{-3}$   & $4.88\times 10^{-2}$  &  $4.89\times 10^{-2}$
\\ \hline
$m_2\,[\mathrm{eV}]$     & $ 8.90 \times 10^{-3}$ & $ 9.10 \times 10^{-3}$  & $4.96 \times 10^{-2}$ &  $4.97\times 10^{-2}$
\\ \hline
$m_3\,[\mathrm{eV}]$     & $4.98 \times 10^{-2}$  & $ 4.99 \times 10^{-2}$  & $2.41\times 10^{-3}$  &  $3.50\times 10^{-3}$
\\ \hline \hline
$\Delta m^2_{21}\,[10^{-5}\mathrm{eV}^2]$       &  $7.60$ &  $7.60$ & $7.50$ & $7.48$
\\ \hline
$|\Delta m^2_{31}|\,[10^{-3}\mathrm{eV}^2]$     &  $2.48$ &  $2.48$ & $2.38$ & $2.38$
\\ \hline
$\sin^2\theta_{12}/10^{-1}$     &  $3.41$  &  $3.41$  & $3.41$ & $3.41$
\\ \hline
$\sin^2\theta_{23}/10^{-1}$     &  $5.67$  &  $5.67$  & $5.73$ & $5.73$
\\ \hline
$\sin^2\theta_{13}/10^{-2}$     &  $2.26$  &  $2.26$  & $2.29$ & $2.29$
\\ \hline
$J_{\text{CP}}/10^{-2}$         &  $-2.71$ &  $-2.71$ & $-2.58$ & $-2.57$
\\ \hline
 \hline
\end{tabular}
\caption{\label{NuMass} Neutrino masses and oscillation parameters
associated to the four chosen benchmark points. }
\end{table}


\section{quark sector}
\label{sec:quark sector}

\begin{table} [ht]
\centering
\begin{tabular}{|c||c|c|c||c|c|c||c|c|c|}
\hline \hline
Field      &  $\Psi_{U}$     &   $\Psi_{C}$    &    $\Psi_{T}$ &  $\Psi_u$  &  $\Psi_c$  & $\Psi_t$ &
$\Psi_d$ & $\Psi_s$ &  $\Psi_b$      \\ \hline
$\Delta(27)$  &  $\mathbf{1}_{0,2}$  &   $\mathbf{1}_{0,1}$ &  $\mathbf{1}_{0,0}$ &
                 $\mathbf{1}_{0,2}$  &   $\mathbf{1}_{0,0}$ &  $\mathbf{1}_{0,2}$ &
                 $\mathbf{1}_{0,1}$  &   $\mathbf{1}_{0,0}$ &  $\mathbf{1}_{0,1}$      \\ \hline
$Z_4$         &   $-i$         &       $-i$       &      $-i$            &
                  $1$          &       $1$        &      $-i$            &
                  $1$          &       $-i$       &      $-i$            \\ \hline
$Z_4^\prime$         &   $1$         &      $1$        &       $1$        &
                  $-1$        &      $-1$       &       $-1$       &
                  $-1$        &      $-1$       &       $-1$                           \\ \hline \hline
\end{tabular}
\caption{\label{tab:tq} Particle content and transformation
  properties of the quark sector under the flavor symmetry $\Delta(27) \otimes  Z_4\otimes Z_4^\prime$.}
\end{table}

The quark transformation properties under the family group $\Delta(27)
\otimes Z_4\otimes Z_4^\prime$ are given in Table~\ref{tab:tq}. At
leading order, the most general invariant Yukawa interactions can be
written as
\begin{eqnarray}
{\cal L}_Y^{q}
&=&
\frac{\sqrt{G}}{ (\Lambda^\prime)^{\frac{5}{2}} }
\Bigg\{
y_{uu}    \sigma_2^\ast  \overline{\Psi}_{U}   \widetilde{H} \Psi_u    +
y_{ct}    \sigma_1^\ast  \overline{\Psi}_{C}   \widetilde{H} \Psi_t    +
y_{tc}    \sigma_2^\ast  \overline{\Psi}_{T}   \widetilde{H} \Psi_c    +
y_{tt}    \sigma_1       \overline{\Psi}_{T}   \widetilde{H} \Psi_t
\nonumber\\
&&\qquad+
y_{ds}    \sigma_1^\ast  \overline{\Psi}_{U}   H \Psi_s    +
y_{db}    \sigma_1       \overline{\Psi}_{U}   H \Psi_b    +
y_{sd}    \sigma_2^\ast  \overline{\Psi}_{C}   H \Psi_d
\nonumber\\
&&\qquad+
y_{ss}    \sigma_1       \overline{\Psi}_{C}   H \Psi_s    +
y_{bb}    \sigma_1^\ast  \overline{\Psi}_{T}   H \Psi_b  \Bigg\}\delta(y-L)
+\text{h.c.}
\label{Eq:Yukawa_quark}
\end{eqnarray}	
Again, after spontaneous electroweak and flavor symmetry breaking, the
mass matrices for the up and down quark sectors read
\begin{eqnarray}
m^{u}
=
\frac{1}{(L\,\Lambda^\prime)^\frac{3}{2}}
\frac{v}{\sqrt{2}}
\left(\begin{array}{ccc}
\widetilde{y}_{uu} v_{\sigma_2}^\ast/\Lambda^\prime&0&0\\
0&0&\widetilde{y}_{ct}v_{\sigma_1}^\ast/\Lambda^\prime\\
0&\widetilde{y}_{tc}v_{\sigma_2}^\ast/\Lambda^\prime&\widetilde{y}_{tt}v_{\sigma_1}/\Lambda^\prime
\end{array}\right)\,,
\nn\\
m^{d}
=
\frac{1}{(L\,\Lambda^\prime)^\frac{3}{2}}
\frac{v}{\sqrt{2}}
\left
(\begin{array}{ccc}
0&\widetilde{y}_{ds}v_{\sigma_1}^\ast/\Lambda^\prime&\widetilde{y}_{db}v_{\sigma_1}/\Lambda^\prime\\
\widetilde{y}_{sd}v_{\sigma_2}^\ast/\Lambda^\prime&\widetilde{y}_{ss}v_{\sigma_1}/\Lambda^\prime&0\\
0&0&\widetilde{y}_{bb}v_{\sigma_1}^\ast/\Lambda^\prime
\end{array}\right)\,.
\end{eqnarray}
where
\begin{eqnarray}
\widetilde{y}_{u_i u_j} &=& y_{u_i u_j} F(L,c_{Q_i},c_{u_j}) \,,
\nn\\
\widetilde{y}_{d_i d_j} &=& y_{d_i d_j} F(L,c_{Q_i},c_{d_j}) \,.
\label{yud}
\end{eqnarray}
The up-type quark mass matrix is already block-diagonal. The
diagonalization of the down-type mass matrix $m^d$ requires a more
careful treatment. For the sake of simplicity, in the following
analysis we denote the $ij$ element of $m^u$ ($m^d$) as $m^u_{ij}$
($m^d_{ij}$). The product of the down-type mass matrix and its adjoint
\begin{equation}
m^d m^{d\dagger}
=
\left(\begin{array}{ccc}
|m^d_{12}|^2+|m^d_{13}|^2  &  m^d_{12} m^{d\ast}_{22}  & m^d_{13} m^{d\ast}_{33}  \\
m^{d\ast}_{12} m^{d}_{22}  & |m^d_{21}|^2+|m^d_{22}|^2 &         0                \\
m^{d\ast}_{13} m^d_{33}    &          0                & |m^d_{33}|^2
\end{array}\right)\,
\end{equation}
can be diagonalized in two steps: in first place, an approximate block
diagonalization
\begin{equation}
U^{d\prime\dagger} m^d m^{d\dagger} U^{d\prime}
\simeq
\left(\begin{array}{ccc}
|m^d_{12}|^2               &   m^d_{12} m^{d\ast}_{22}   &          0                  \\
m^{d\ast}_{12} m^{d}_{22}  &  |m^d_{21}|^2+|m^d_{22}|^2  &          0                  \\
0                          &           0                 & |m^d_{33}|^2
\end{array}\right)\,,
\end{equation}
is accomplished with the aid of the transformation matrix
\begin{equation}
U^{d\prime}
\simeq
\left(\begin{array}{ccc}
  1               &   0  &   \epsilon   \\
  0               &   1  &   0   \\
 -\epsilon^\ast &   0  &   1
\end{array}\right)\,,
\end{equation}
and subsequently the diagonalization is completed through a unitary
rotation of the upper block.  This approximation is consistent
provided $|m^d_{33}|>>|m^d_{12}|,|m^d_{13}|,|m^d_{22}|$ and
$|\epsilon|<<1$.
The resulting diagonalization matrices for the up and down sectors can
be parameterized as
\begin{eqnarray}
U_{u}
&=&
\left(\begin{array}{ccc}
1&0&0\\
0&\cos{\theta_u}                         &\sin{\theta_u} e^{i\varphi_u}          \\
0&-\sin{\theta_u} e^{-i\varphi_u}        &\cos{\theta_u}
\end{array}\right)\,,\\
U_{d}
&\simeq&
\left(\begin{array}{ccc}
\cos{\theta_d}                     &\sin{\theta_d} e^{i\varphi_d}             &    \epsilon \\
-\sin{\theta_d} e^{-i\varphi_d}    &\cos{\theta_d}                            &   0 \\
-\epsilon^*\cos{\theta_d}          &-\epsilon^*\sin{\theta_d} e^{i\varphi_d}  &    1
\end{array}\right)\,,\nn
\end{eqnarray}
with
\begin{equation}
\begin{array}{ccccc}
\tan{2\theta_u}=2|Z_u|/X^{-}_u\,,  &   \qquad & \varphi_u=\arg Z_u\,, & \qquad &\\
\tan{2\theta_d}=2|Z_d|/X^{-}_d\,,  &   \qquad & \varphi_d=\arg Z_d\,, & \qquad & \epsilon = B_d/A_d\,,
\end{array}
\label{par_quarks}
\end{equation}
and
\begin{equation}
\begin{array}{lcccc}
X^{\pm}_u=|m^u_{33}|^2+|m^u_{32}|^2\pm |m^u_{23}|^2\,, &\qquad&
Y_u=m^u_{23} m^{u*}_{32}\,, &\qquad& Z_u= m^u_{23} m^{u*}_{33}\,,\\
X_d^{\pm}=|m^d_{22}|^2+|m^d_{21}|^2\pm |m^d_{12}|^2\,, &\qquad&
Y_d=m^d_{12}m^{d*}_{21}\,,  &\qquad& Z_d= m^d_{12} m^{d*}_{22}\,,\\
\,\, A_d=|m^d_{33}|^2-|m^d_{12}|^2-|m^d_{13}|^2\,,&\qquad& B_d=m^d_{13} m^{d*}_{33}\,. &\qquad&
\label{aux_quarks}
\end{array}
\end{equation}
Correspondingly, the quark mass eigenvalues can be expressed in terms of $M^{\pm}$, defined in Eq.~\eqref{Mpm}, as
\begin{equation}
\begin{array}{lclcl}
m_u=|m^u_{11}|\,, & \qquad & m_c=\frac{1}{\sqrt{2}}M^-\big(X^{+}_u,Y_u\big)\,, & \qquad &
m_t=\frac{1}{\sqrt{2}}M^+\big(X^{+}_u,Y_u\big)\,,\\
m_d=\frac{1}{\sqrt{2}}M^-\big(X^{+}_d,Y_d\big)\,,  & \qquad & m_s=\frac{1}{\sqrt{2}}M^+\big(X^{+}_d,Y_d\big)\,,
& \qquad & m_b=|m^b_{33}|\,,
\end{array}
\end{equation}
so that the CKM matrix is given by
\begin{eqnarray}
V_{\text{CKM}}
&=&U_{u}^\dagger  U_{d}
\\
&\simeq&
\left(
\begin{array}{ccc}
 \cos \theta _d &
 e^{i\varphi _d} \sin \theta_d &
 \epsilon
\\
 -e^{-i \varphi _d} \cos \theta_u \sin \theta_d - e^{i \varphi_u} \sin \theta_u \cos \theta_d \epsilon^*&
 \cos \theta_d \cos \theta_u  - e^{i (\varphi_u + \varphi_d ) } \sin \theta_u \sin \theta_d \epsilon^* &
 -e^{i \varphi _u} \sin \theta _u
\\
 -e^{-i \left(\varphi _d+\varphi_u\right)} \sin \theta_d \sin \theta_u - \cos \theta_u \cos \theta_d \epsilon^*    &
 e^{-i \varphi _u} \cos \theta_d \sin\theta_u -  e^{i \varphi_d} \cos \theta_u \sin \theta_d \epsilon^*  &
 \cos\theta _u
\\
\end{array}
\right)\,.
\nn
\end{eqnarray}
Hence, the quark sector Dirac CP phase (in PDG convention) and the
Jarlskog invariant take the form
\begin{eqnarray}
\delta^q_{\text{CP}}&=&\pi-\arg(\epsilon)+\varphi _d+\varphi_u\,,\label{deltaq}\\
J^{q}_{\text{CP}}&\simeq & \frac{1}{4} |\epsilon| \sin 2 \theta _d \sin 2 \theta _u \sin \delta^q_{\text{CP}}\label{Jq}\,.
\end{eqnarray}
According to Eq.~(\ref{yud}), the size of up and down mass matrix
elements is determined by the overlap of the 5D quark field zero mode
profiles, {\it i.e.}, $m^u_{ij}\propto f_L^{(0)}(L,c_{Q_i})
f_R^{(0)}(L,c_{u_j})$ and $m^d_{ij}\propto f_L^{(0)}(L,c_{Q_i})
f_R^{(0)}(L,c_{d_j})$. If the wave function localization parameters
$c_{Q_i}$, $c_{u_i}$, $c_{d_i}$ are chosen such that the quark zero
mode profiles obey
\begin{eqnarray}
&&f_L^{(0)}(L,c_{U})\ll f_L^{(0)}(L,c_{C})\ll f_L^{(0)}(L,c_{T})\,,\nn\\
&&f_R^{(0)}(L,c_{u})\ll f_R^{(0)}(L,c_{c})\ll f_R^{(0)}(L,c_{t})\,,\nn\\
&&f_R^{(0)}(L,c_{d})\ll f_R^{(0)}(L,c_{s})\ll f_R^{(0)}(L,c_{b})\,,
\end{eqnarray}
then the elements of $m^u$ and $m^d$ approximately satisfy
\begin{eqnarray}
m^u_{11} \ll m^u_{23}\sim m^u_{32} \ll m^u_{33}\,, \hspace{0.5cm}
m^d_{12}\sim m^d_{21} \ll m^d_{22} \ll m^d_{33}\,, \hspace{0.5cm} m^d_{13} \ll m^d_{33}\,,
\end{eqnarray}
justifying the perturbative diagonalization performed on $m^dm^{d\dagger}$.  These relations imply that $X^{+}_{u,d}\gg |Y_{u,d}|$ holds, and therefore, a rough estimate for the mixing parameters and quark mass spectrum is
\begin{equation}
\begin{array}{ccccc}
\theta_{u}\sim \left| \frac{m^u_{23}}{m^u_{33}} \right| \sim \frac{f_L^{(0)}(L,c_{C})}{f_L^{(0)}(L,c_{T})}\,,&\qquad&
\theta_{d}\sim \left| \frac{m^d_{12}}{m^d_{22}} \right| \sim \frac{f_L^{(0)}(L,c_{U})}{f_L^{(0)}(L,c_{C})}\,,&\qquad&
|\epsilon|    \sim \left| \frac{m^d_{13}}{m^d_{33}} \right| \sim \frac{f_L^{(0)}(L,c_{U})}{f_L^{(0)}(L,c_{T})}\,,
\\
m_u \sim  \left|m^u_{11}\right|\,,&\qquad&
m_c \sim  \left|\frac{m^u_{23} m^u_{32} }{m^u_{33} } \right|\,,&\qquad&
m_t \sim  \left|m^u_{33}\right|\,,\hspace{0.4cm}
\\
m_d \sim  \left|\frac{m^d_{12} m^d_{21} }{m^d_{22} } \right|\,,&\qquad&
m_s \sim  \left|m^d_{22}\right|\,,&\qquad&
m_b \sim  \left|m^d_{33}\right|\,.
\end{array}
\end{equation}
Thus, in order to reproduce plausible quark masses and mixings, namely:
\begin{equation}
\begin{array}{c}
\theta_{u}\sim 10^{-1}\,,\qquad
\theta_{d}\sim 10^{-2}\,,\qquad
|\epsilon|  \sim 10^{-3}\,,
\\
m_u:m_c:m_t \sim 10^{-5}:10^{-2}:1\,,
\\
m_d :m_s :m_b \sim 10^{-3}:10^{-2}:1\,,
\end{array}
\end{equation}
the quark zero mode profiles must observe the following hierarchy:
\begin{eqnarray}
&&f_L^{(0)}(L,c_{U}): f_L^{(0)}(L,c_{C}): f_L^{(0)}(L,c_{T})\sim 10^{-3}: 10^{-1}:1 \,,\nn\\
&&f_R^{(0)}(L,c_{u}): f_R^{(0)}(L,c_{c}): f_R^{(0)}(L,c_{t})\sim 10^{-2}: 10^{-1}:1 \,,\nn\\
&&f_R^{(0)}(L,c_{d}): f_R^{(0)}(L,c_{s}): f_R^{(0)}(L,c_{b})\sim 10^{-1}: 10^{-1}:1 \,.
\end{eqnarray}

To conclude this section, an explicit realization of quark masses and mixings is presented. The choice
$c_U = 1.97$,
$c_C = 1.92$,
$c_T = 1.83$,
$c_u = -0.76$,
$c_c = -0.62$,
$c_t = -0.56$,
$c_d = -0.74$,
$c_s = -0.69$,
$c_b = -0.68$,
$y_{uu} =-0.438 - 0.954 i$,
$y_{ct} =-0.360 - 1.038 i$,
$y_{tc} = 1.147 - 0.273 i$,
$y_{tt} =-0.372 - 1.073 i$,
$y_{ds} =-0.966 - 0.285 i$,
$y_{db} = 0.290 + 0.400 i$,
$y_{sd} = 0.838 - 0.226 i$,
$y_{ss} =-0.703 - 0.207 i$,
$y_{bb} = 0.637 - 0.879 i$,
generates the quark mass spectrum
\begin{equation}
\begin{array}{lllll}
m_u= 2.30 \, \text{MeV}\,, &\qquad &
m_c= 1.275\,\text{GeV}\,,   &\qquad &
m_t= 173  \,\text{GeV}\,,\\
m_d= 4.80\,\text{MeV}\,,   &\qquad &
m_s= 95.0\,\text{MeV}\,,   &\qquad &
m_b= 4.18\,\text{GeV}\,,
\end{array}
\end{equation}
and fixes the magnitude of $V_{CKM}$ elements at
\begin{equation}
|V_{\text{CKM}}|
=
\left(
\begin{array}{ccc}
 0.974 & 0.225 & 0.0035 \\
 0.225 & 0.973 & 0.0414 \\
 0.0089 & 0.041 & 0.999 \\
\end{array}
\right)
\,.
\end{equation}
Finally, the obtained values for the Dirac CP phase and the Jarlskog invariant are
\begin{equation}
\delta^{q}_{\text{CP}}=1.25 \,,\qquad
J^{q}_{\text{CP}}
=3.06\times 10^{-5}
\,.
\end{equation}
The resulting quark masses and mixings are consistent with the current
experimental data~\cite{Agashe:2014kda}, and the precision of the
results can be improved by incorporating high order corrections,
addressed in the next section.

\section{\label{sec:NLO}High Order Corrections}

From the particle content and above transformation properties, one
finds that nontrivial high order corrections to the charged lepton
sector are absent in the present model. The next-to-leading order
(NLO) corrections to the neutrino Yukawa interactions are given by
\begin{eqnarray}
{\cal \delta L}^{\nu}_Y &=&
\sqrt{G}
\frac{x_2 }{ (\Lambda^\prime)^{\frac{9}{2}} }
\big[(\xi^\ast \xi^\ast)_{\mathbf{3}} \sigma^\ast_2   \overline{\Psi}_l\big]_{\mathbf{1}_{0,0}}  \widetilde{H} \Psi_{\nu_2}
\delta(y-L)
+
\text{h.c.}
\end{eqnarray}
However, the contribution of these terms to the neutrino masses and
mixing parameters can be absorbed by a proper redefinition of the
parameter $y_{22}$ after SSB. Hence, in order to estimate the effects
of higher order corrections in this sector, we need to investigate the
Yukawa terms involving an additional
$(v_{\text{IR}}/\Lambda^{\prime})^2$ suppression with respect to the
lowest order terms in Eq.~\eqref{eq:wl_LO}, where we have introduced
$v_{\text{IR}}$ to characterize the magnitude of $v_\xi\sim
v_{\sigma_1}\sim v_{\sigma_2}$.

The contraction of the field products
$\overline{\Psi}_l\widetilde{H}\Psi_{\nu_1}$,
$\overline{\Psi}_l\widetilde{H}\Psi_{\nu_3}$, transforming as
$(\mathbf{\overline{3}},-1,-1)$ under $\Delta(27) \otimes Z_4\otimes
Z_4^\prime$, with the flavon operators
\begin{eqnarray}
&&\frac{1}{ (\Lambda^\prime)^{\frac{11}{2}}} (\xi\xi^\ast)_{\mathbf{1}_{a,2}}\xi\sigma_1\,,      \hspace{1cm}
\frac{1}{ (\Lambda^\prime)^{\frac{11}{2}}} (\xi\xi^\ast)_{\mathbf{1}_{a,1}}\xi\sigma_1^\ast\,,   \hspace{1cm}
\frac{1}{ (\Lambda^\prime)^{\frac{11}{2}}} \xi\sigma_1^3\,,                                    \hspace{1cm}
\frac{1}{ (\Lambda^\prime)^{\frac{11}{2}}} \xi\sigma_1^{\ast3}\,,                                \hspace{1cm}
\end{eqnarray}
as well as the combination of
$\overline{\Psi}_l\widetilde{H}\Psi_{\nu_2}\sim(\mathbf{\overline{3}},i,-1)$ and
\begin{eqnarray}
\frac{1}{(\Lambda^\prime)^{\frac{11}{2}}} (\xi\xi^\ast)_{\mathbf{1}_{a,b}}\xi\sigma_2\,,         \hspace{1cm}
\frac{1}{(\Lambda^\prime)^{\frac{11}{2}}} \xi\sigma_1^2        \sigma_2\,,                       \hspace{1cm}
\frac{1}{(\Lambda^\prime)^{\frac{11}{2}}} \xi\sigma_1^{\ast2}  \sigma_2\,,                       \hspace{1cm}
\end{eqnarray}
provide the desired high order corrections to the neutrino Yukawa interactions. In the above expressions, the indices $a,b=0,1,2$ label the different singlets of $\Delta(27)$. Additional terms that can be absorbed into $y_{11}$, $y_{13}$, $y_{22}$, $y_{31}$ and $y_{33}$ have been omitted. Taking into consideration these corrections, the neutrino mass matrix $m_\nu$ can be roughly written as
\begin{eqnarray}
m_\nu \simeq
\frac{1}{ (L\,\Lambda^\prime)^\frac{3}{2} }
\frac{v}{\sqrt{2}}
\left[
\frac{v_\xi}{\Lambda^\prime}
\left(
\begin{array}{ccc}
\widetilde{y}_{11} \frac{v_{\sigma_1}}{\Lambda^\prime}  &  0  &  \widetilde{y}_{13} \frac{v_{\sigma_1}}{\Lambda^\prime}  \\
0                                                       &  \widetilde{y}_{22} \frac{v_{\sigma_2}}{\Lambda^\prime}  &  0  \\
\widetilde{y}_{31} \frac{v_{\sigma_1}^\ast}{\Lambda^\prime}  &  0  &  \widetilde{y}_{33} \frac{v_{\sigma_1}^\ast}{\Lambda^\prime}
\end{array}
\right)
+
\left(\frac{v_{\text{IR}}}{\Lambda^\prime}\right)^4
\left(
\begin{array}{ccc}
0                  & \widetilde{x}_{12}  &               0      \\
\widetilde{x}_{21} &              0      &  \widetilde{x}_{23}  \\
0                  & \widetilde{x}_{32}  &               0
\end{array}
\right)
\right]\,,
\end{eqnarray}
with $\widetilde{x}_{ij}=x_{ij}F(L,c_l,c_{\nu_j})$, and $x_{ij}$ as
dimensionless parameters of order $O(1)$.

Working under the same numerical framework established in Section
\ref{lepton}, one can readily estimate the shift in the neutrino
oscillation parameters induced by high order corrections of the Yukawa
interaction.  Particularly, in Case I, taking $x_{ij}$ as random
complex numbers with magnitudes ranging from 2 to 6, and
$v_{\text{IR}}=0.1$, the resulting deviations in the neutrino mixing
parameters with respect to their LO values can be estimated
as
\begin{eqnarray}
\delta s^2_{12}    \sim  0.01\,   \hspace{1cm}
\delta s^2_{23}    \sim  0.01\,   \hspace{1cm}
\delta s^2_{13}    \sim  0.001\,  \hspace{1cm}
\delta J_{CP} \sim  0.001\,.
\label{NPDev}
\end{eqnarray}
On the other hand, the corrections to the neutrino mass splittings are
negligible
\begin{eqnarray}
\delta \left(\Delta m^2_{21}\right)    \sim  10^{-7}\,\mathrm{eV^2} \,,  \hspace{1cm}
\delta \left(\Delta |m^2_{31}|\right)    \sim  10^{-6}\,\mathrm{eV^2} \,.   \hspace{1cm}
\end{eqnarray}
From Eq.~\eqref{NPDev}, it is clear that high order corrections can
easily drive $s^2_{12}$ into its $1\sigma$ region while keeping the
remaining parameters optimal.

Turning to the quark sector, every bilinear formed by
$\overline{\Psi}_{Q_i}$ and $\Psi_{u_i}$ or $\Psi_{d_i}$ can produce a
high order correction to the Yukawa interaction whenever it is
contracted with the adequate cubic flavon operator.  Beside terms that
can be absorbed by a redefinition of $y_{u_iu_j}$ or $y_{d_id_j}$ in
Eq.~\eqref{Eq:Yukawa_quark}, all the NLO contributions can be classified
into three categories:
\begin{itemize}
\item Invariant products of $\overline{\Psi}_{U}   \widetilde{H} \Psi_c$, $\overline{\Psi}_{C}   \widetilde{H} \Psi_u$, $\overline{\Psi}_{T}   H \Psi_d \, \sim (\mathbf{1}_{0,1},i,-1)$ with
\begin{eqnarray}
\frac{1}{(\Lambda^\prime)^{\frac{9}{2}}}(\xi\xi^\ast)_{\mathbf{1}_{0,2}}   \sigma_2^\ast\,,             \hspace{2cm}
\frac{1}{(\Lambda^\prime)^{\frac{9}{2}}}\sigma_1^2\sigma_2^{\ast}\,.                                \hspace{1cm}
\end{eqnarray}
\item Invariant products of $\overline{\Psi}_{U}   \widetilde{H} \Psi_t$, $\overline{\Psi}_{C}   H \Psi_b$, $\overline{\Psi}_{T}   H \Psi_s\,\sim (\mathbf{1}_{0,0},1,-1)$ with
\begin{eqnarray}
\frac{1}{(\Lambda^\prime)^{\frac{9}{2}}}(\xi\xi^\ast)_{\mathbf{1}_{0,2}} \sigma_1\,,               \hspace{1cm}
\frac{1}{(\Lambda^\prime)^{\frac{9}{2}}}(\xi\xi^\ast)_{\mathbf{1}_{0,1}} \sigma_1^\ast\,,          \hspace{1cm}
\frac{1}{(\Lambda^\prime)^{\frac{9}{2}}}\sigma_1^3\,,                                              \hspace{1cm}
\frac{1}{(\Lambda^\prime)^{\frac{9}{2}}}\sigma_1^{\ast3}\,.                                        \hspace{0.3cm}
\end{eqnarray}
\item Invariant products of $\overline{\Psi}_{C}   \widetilde{H} \Psi_c$, $\overline{\Psi}_{T}   \widetilde{H} \Psi_u$ and $\overline{\Psi}_{U}   H \Psi_d\, \sim (\mathbf{1}_{0,2},i,-1)$ with
\begin{eqnarray}
\frac{1}{(\Lambda^\prime)^{\frac{9}{2}}}(\xi\xi^\ast)_{\mathbf{1}_{0,1}}   \sigma_2^\ast\,,        \hspace{2cm}
\frac{1}{(\Lambda^\prime)^{\frac{9}{2}}}\sigma_1^{\ast2}\sigma_2^{\ast}\,.                                      \hspace{1cm}
\end{eqnarray}
\end{itemize}

Again, after symmetry breaking, the quark mass matrices $m_u$ and
$m_d$ can be approximately written as
\begin{eqnarray}
m^{u}
&=&
\frac{1}{(L\,\Lambda^\prime)^\frac{3}{2}}
\frac{v}{\sqrt{2}}
\left[
\left(\begin{array}{ccc}
\widetilde{y}_{uu} v_{\sigma_2}^\ast/\Lambda^\prime&0&0\\
0&0&\widetilde{y}_{ct}v_{\sigma_1}^\ast/\Lambda^\prime\\
0&\widetilde{y}_{tc}v_{\sigma_2}^\ast/\Lambda^\prime&\widetilde{y}_{tt}v_{\sigma_1}/\Lambda^\prime
\end{array}\right)
+
\left(\frac{v_{\text{IR}}}{\Lambda^\prime}\right)^3
\left(\begin{array}{ccc}
0&\widetilde{x}_{uc} &\widetilde{x}_{ut}\\
\widetilde{x}_{cu}&\widetilde{x}_{cc}&0\\
\widetilde{x}_{tu}&0&0
\end{array}\right)
\right]
\,,
\nn\\
m^{d}
&=&
\frac{1}{(L\,\Lambda^\prime)^\frac{3}{2}}
\frac{v}{\sqrt{2}}
\left[
\left(\begin{array}{ccc}
0&\widetilde{y}_{ds}v_{\sigma_1}^\ast/\Lambda^\prime&\widetilde{y}_{db}v_{\sigma_1}/\Lambda^\prime\\
\widetilde{y}_{sd}v_{\sigma_2}^\ast/\Lambda^\prime&\widetilde{y}_{ss}v_{\sigma_1}/\Lambda^\prime&0\\
0&0&\widetilde{y}_{bb}v_{\sigma_1}/\Lambda^\prime
\end{array}\right)
+
\left(\frac{v_{\text{IR}}}{\Lambda^\prime}\right)^3
\left(\begin{array}{ccc}
\widetilde{x}_{dd}&0&0\\
0&0&\widetilde{x}_{sb}\\
\widetilde{x}_{bd}&\widetilde{x}_{bs}&0
\end{array}\right)
\right]
\,.
\end{eqnarray}
Here we have defined
$\widetilde{x}_{u_iu_j}=x_{u_iu_j}F(L,c_{Q_i},c_{u_j})$ and
$\widetilde{x}_{d_id_j}=x_{d_id_j}F(L,c_{Q_i},c_{d_j})$, where the
couplings $x_{u_iu_j}$ and $x_{d_id_j}$ represent dimensionless
parameters of order $O(1)$. As a numerical example, taking
$x_{u_iu_j}$, $x_{d_id_j}$ as random complex numbers with magnitudes
ranging from 1 to 4 for $x_{uc}$, $x_{cu}$, $x_{bd}$, $x_{cc}$,
$x_{tu}$, $x_{dd}$, and from 2 to 6 for $x_{ut}$, $x_{sb}$, $x_{bs}$,
while keeping the values of $c_{Q_i}$, $c_{u_i}$, $c_{d_i}$,
$y_{u_iu_j}$ and $y_{d_id_j}$ reported in Section \ref{sec:quark
  sector}, the order of deviation with respect to the LO
values of the quark masses is
\begin{equation}
\begin{array}{lllll}
\delta m_u \sim 0.001\,  \text{MeV}\,, &\qquad &
\delta m_c \sim 10   \,  \text{MeV}\,, &\qquad &
\delta m_t \sim 0.1  \,  \text{MeV}\,, \\
\delta m_d \sim 0.1  \,  \text{MeV}\,, &\qquad &
\delta m_s \sim 0.1  \,  \text{MeV}\,, &\qquad &
\delta m_b\sim 0.5\text{MeV}\,.
\end{array}
\end{equation}
The corresponding correction to the first order CKM matrix is of order
\begin{eqnarray}
\delta|V_{\text{CKM}}|\sim
\left(
\begin{array}{ccc}
 0.001  & 0.005 & 0.0001 \\
 0.005  & 0.001 & 0.001  \\
 0.0005 & 0.001 & 0.00005 \\
\end{array}
\right)\,,
\end{eqnarray}
and the values for the quark CP violating phase and the Jarlskog
invariant are displaced by
\begin{equation}
\delta (\delta^{q}_{\text{CP}})\sim 0.1 \,,\qquad
\delta J^{q}_{\text{CP}}
\sim  10^{-6}
\,.
\end{equation}
As for the lepton sector, it is not difficult to find parameter values
reproducing the quark mass and mixing parameters required to fit the
current experimentally observed values.

\section{\label{sec:conclusions} Conclusions}

\begin{figure}[t!]
\centering
\includegraphics[scale=0.75]{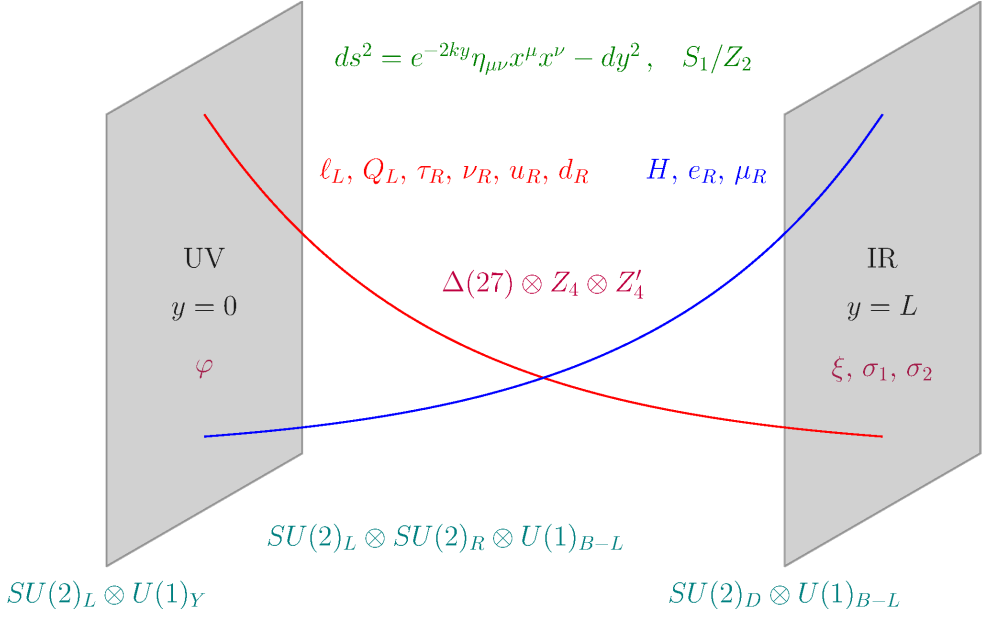}
\caption{Pictorial description of the basic warped model structure, showing the UV (IR) peaked nature of the \sm fields.}
\label{fig:cartoon}
\end{figure}

We have proposed a five-dimensional warped model in which all \sm
fields propagate into the bulk. Its structure is
summarized in the ``cartoon'' depicted in Figure~\ref{fig:cartoon}.
Mass hierarchies in principle arise from an adequate choice of the
bulk shape parameters, while fermion mixing angles are constrained by
relations which follow from the postulated $\Delta(27)$ flavor
symmetry group, broken on the branes by a set of flavon fields.
The neutrino mixing parameters and the Dirac CP violation phase are
described in terms of just two independent parameters at leading order. This leads to stringent predictions for the lepton mixing
matrix which should be tested in future neutrino oscillation
experiments.
Likewise the scheme also includes the quark sector, providing an
adequate description of the quark mixing matrix. The effect of
next-to-leading order contributions is estimated to be fully
consistent with the experimental requirements.

\section*{Acknowledgements}
This work is supported by the National Natural Science Foundation of China under Grant Nos. 11275188, 11179007 and 11522546; by the Spanish grants FPA2014-58183-P, Multidark CSD2009-00064, SEV-2014-0398 (MINECO) and PROMETEOII/2014/084 (Generalitat Valenciana). A.D.R. and C.A.V-A. acknowledge support from CONACyT (Mexico), grants 250610 and 251357.

\newpage
\begin{appendix}

  \section{\label{sec:App_A}Group theory of $\Delta(27)$
and its  representation}
\setcounter{equation}{0}

The $\Delta(27)$ group is isomorphic to $(Z_3\otimes Z_3)\rtimes
Z_3$. It can be conveniently expressed in terms of three generators
$a$, $a^{\prime}$ and $b$ which satisfy the following relations:
\begin{eqnarray}
&&a^3={a^\prime}^3=b^3=1,\qquad aa^{\prime}=a^{\prime}a,\nonumber\\
&&bab^{-1}=a^{-1}{a^\prime}^{-1},\qquad ba^{\prime}b^{-1}=a\,.
\end{eqnarray}
All $\Delta(27)$ elements can be written into the form
$b^{k}a^{m}{a^{\prime}}^{n}$, with $k, m, n=0, 1, 2$. The group has 11
conjugacy classes, given by
\begin{eqnarray}
\nonumber&&1C_1=\{ 1  \},       \\
\nonumber&&1C_1^{(1)}=\{a{a^\prime}^2  \},     \\
\nonumber&&1C_1^{(2)}=\{ a^2a^\prime  \},    \\
\nonumber&&3C_3^{(0,1)}=\{a, a^\prime, a^2 {a^\prime}^2 \},   \\
\nonumber&&3C_3^{(0,2)}=\{{a^2, a^\prime}^2, aa^\prime\},    \\
\nonumber&&3C_3^{(1,0)}=\{b, ba{a^\prime}^2, b a^2a^\prime\},    \\
\nonumber&&3C_3^{(1,1)}=\{ba, ba^\prime, ba^2{a^\prime}^2 \},    \\
\nonumber&&3C_3^{(1,2)}=\{ba^2, baa^\prime, b{a^\prime}^2\},    \\
\nonumber&&3C_3^{(2,0)}=\{b^2, b^2a{a^\prime}^2, b^2a^2a^\prime \},    \\
\nonumber&&3C_3^{(2,1)}=\{b^2a, b^2a^{\prime}, b^2a^2{a^\prime}^2 \},    \\
&&3C_3^{(2,2)}=\{b^2a^2, b^2aa^{\prime}, b^2{a^\prime}^2\}\,.
\end{eqnarray}
\begin{table}[t!]
\begin{center}
\begin{tabular}{|c|c|c|c|c|c|c|c|c|c|c|c|}\hline\hline
& ~$\chi_{\mathbf{1}_{0,0}}$~
& ~$\chi_{\mathbf{1}_{0,1}}$~
& ~$\chi_{\mathbf{1}_{0,2}}$~
& ~$\chi_{\mathbf{1}_{1,0}}$~
& ~$\chi_{\mathbf{1}_{1,1}}$~
& ~$\chi_{\mathbf{1}_{1,2}}$~
& ~$\chi_{\mathbf{1}_{2,0}}$~
& ~$\chi_{\mathbf{1}_{2,1}}$~
& ~$\chi_{\mathbf{1}_{2,2}}$~
& ~~$\chi_{\mathbf{3}}$~~
& ~~$\chi_{\mathbf{\overline{3}}}$~~\\\hline
$1C_1$&1&1         &1         &1         &1         &1         &1         &1         &1         &3          &3          \\\hline
$1C_1^{(1)}$&1&1         &1         &1         &1         &1         &1         &1         &1         &$3\omega^2$&$3\omega$  \\\hline
$1C_1^{(2)}$&1&1         &1         &1         &1         &1         &1         &1         &1         &$3\omega$  &$3\omega^2$\\\hline
$3C_3^{(0,1)}$&1&$\omega$  &$\omega^2$&1         &$\omega$  &$\omega^2$&1         &$\omega$  &$\omega^2$&0          &0          \\\hline
$3C_3^{(0,2)}$&1&$\omega^2$&$\omega$  &1         &$\omega^2$&$\omega$  &1         &$\omega^2$&$\omega$  &0          &0          \\\hline
$3C_3^{(1,0)}$&1&1         &1         &$\omega$  &$\omega$  &$\omega$  &$\omega^2$&$\omega^2$&$\omega^2$&0          &0          \\\hline
$3C_3^{(1,1)}$&1&$\omega$  &$\omega^2$&$\omega$  &$\omega^2$&1         &$\omega^2$&1         &$\omega$  &0          &0          \\\hline
$3C_3^{(1,2)}$&1&$\omega^2$&$\omega$  &$\omega$  &1         &$\omega^2$&$\omega^2$&$\omega$  &1         &0          &0          \\\hline
$3C_3^{(2,0)}$&1&1         &1         &$\omega^2$&$\omega^2$&$\omega^2$&$\omega$  &$\omega$  &$\omega$  &0          &0          \\\hline
$3C_3^{(2,1)}$&1&$\omega$  &$\omega^2$&$\omega^2$&1         &$\omega$  &$\omega$  &$\omega^2$&1         &0          &0          \\\hline
$3C_3^{(2,2)}$&1&$\omega^2$&$\omega$  &$\omega^2$&$\omega$  &1         &$\omega$  &1         &$\omega^2$&0          &0          \\\hline\hline
\end{tabular}
\caption{\label{tab:character}Character table of $\Delta(27)$.}
\end{center}
\end{table}
The $\Delta(27)$ has nine one dimensional representations, which we
denote as $\mathbf{1}_{k,r}\,\, (k,r=0,1,2)$, and two three
dimensional irreducible representations $\mathbf{3}$ and
$\mathbf{\overline{3}}$. The explicit form of the group generators in
each irreducible representation is
\begin{equation}
\begin{array}{lccc}
\mathbf{1}_{k,r}:&~
a=\omega^r,  & ~
a^{\prime}=\omega^r & ~
b=\omega^k,\\
\mathbf{3}:&~
a=\left(
\begin{array}{ccc}
\omega  &     0      &    0      \\
  0     &     1      &    0      \\
  0     &     0      & \omega^2
\end{array}
\right), &~
a^{\prime}=
\left(
\begin{array}{ccc}
\omega^2  &     0      &    0      \\
  0     & \omega   &    0      \\
  0     &     0      &    1
\end{array}
\right),  & ~
b=
\left(
\begin{array}{ccc}
  0     &     1      &    0      \\
  0     &     0      &    1      \\
  1     &     0      &    0
\end{array}
\right),\\
\mathbf{\overline{3}}:&~
a=
\left(
\begin{array}{ccc}
\omega^2&     0      &    0      \\
  0     &     1      &    0      \\
  0     &     0      & \omega
\end{array}
\right),  &~
a^{\prime}=
\left(
\begin{array}{ccc}
\omega &     0      &    0      \\
  0     &  \omega^2   &    0      \\
  0     &     0      &    1
\end{array}
\right), & ~
b=
\left(
\begin{array}{ccc}
  0     &     1      &    0      \\
  0     &     0      &    1      \\
  1     &     0      &    0
\end{array}
\right)\,,
\end{array}
\end{equation}
where $\omega=e^{2\pi i/3}$ is the cube root of unity. Notice that
$\mathbf{3}$ and $\mathbf{\overline{3}}$ are complex representations
dual to each other. From the character table of the group, shown in
Table~\ref{tab:character}, we can straightforwardly obtain the
Kronecker products between the various representations
\begin{eqnarray}
\nonumber&&\mathbf{1}_{k,r}\otimes\mathbf{1}_{k^{\prime},r^{\prime}}=\mathbf{1}_{[k+k^{\prime}],[r+r^{\prime}]}, \qquad  \mathbf{3}\otimes\mathbf{1}_{k,r}=\mathbf{3}, \qquad \mathbf{\overline{3}}\otimes\mathbf{1}_{k,r}=\mathbf{\overline{3}},\\
&&\mathbf{3}\otimes\mathbf{\overline{3}}= \sum_{k,r=0}^2\mathbf{1}_{k,r}, \qquad \mathbf{3}\otimes\mathbf{3}=\mathbf{\overline{3}}\oplus\mathbf{\overline{3}}\oplus\mathbf{\overline{3}},\qquad
\mathbf{\overline{3}}\otimes\mathbf{\overline{3}}=\mathbf{3}\oplus\mathbf{3}\oplus\mathbf{3}\,,
\end{eqnarray}
where $[n]$ stands for $n$ mod $3$, whenever $n$ is an
integer. Starting from the representation matrices of the generators
in different irreducible representations, we can calculate the
Clebsch-Gordan (CG) coefficients for the Kronecker products listed
above. All CG coefficients are presented in the form
$\alpha\otimes\beta$, where $\alpha_{i}$ stands for the elements of the
first representation and $\beta_{j}$ those of the second one. In the
following, we adopt the convention
$\alpha_{[3]}=\alpha_{0}\equiv\alpha_{3}$.
\begin{itemize}

\item{$\mathbf{1}_{k,r}\otimes\mathbf{1}_{k^\prime, r^\prime}=  \mathbf{1}_{[k+k^{\prime}], [r+r^{\prime}]}      $}

\begin{eqnarray*}
\left(
\begin{array}{c}
\alpha_1
\end{array}
\right)_{\mathbf{1}_{k, r}}
\otimes&
\left(
\begin{array}{c}
\beta_1
\end{array}
\right)_{\mathbf{1}_{k^\prime, r^\prime}}
=~
\left(
\begin{array}{c}
\alpha_1  \beta_1
\end{array}
\right)_{ \mathbf{1}_{[k+k^{\prime}], [r+r^{\prime}]}}\,.
\end{eqnarray*}

\item{ $\mathbf{3}\otimes \mathbf{1}_{k,r}  = \mathbf{3} $}

\begin{eqnarray*}
\left(
\begin{array}{c}
\alpha_1
\\
\alpha_2
\\
\alpha_3
\end{array}
\right)_{\mathbf{3}}
\otimes&
\left(
\begin{array}{c}
\beta_1
\end{array}
\right)_{\mathbf{1}_{k,r}}
=~
\left(
\begin{array}{c}
\alpha_{[1+r]}  \beta_1
\\
\omega^k \alpha_{[2+r]}  \beta_1
\\
\omega^{2k} \alpha_{[3+r]}  \beta_1
\end{array}
\right)_{\mathbf{3}}\,.
\end{eqnarray*}

\item{ $\mathbf{\overline{3}}\otimes \mathbf{1}_{k,r}  = \mathbf{\overline{3}} $}

\begin{eqnarray*}
\left(
\begin{array}{c}
\alpha_1
\\
\alpha_2
\\
\alpha_3
\end{array}
\right)_{\mathbf{\overline{3}}}
\otimes&
\left(
\begin{array}{c}
\beta_1
\end{array}
\right)_{\mathbf{1}_{k,r}}
=~~
\left(
\begin{array}{c}
\alpha_{[1-r]}  \beta_1
\\
\omega^k \alpha_{[2-r]}  \beta_1
\\
\omega^{2k} \alpha_{[3-r]}  \beta_1
\end{array}
\right)_{\mathbf{\overline{3}}}\,.
\end{eqnarray*}

\item{$\mathbf{3}\otimes\mathbf{\overline{3}}= \sum_{k,r=0}^2 \mathbf{1}_{k,r} $}

\begin{eqnarray*}
&&\left(
\begin{array}{c}
\alpha_1
\\
\alpha_2
\\
\alpha_3
\end{array}
\right)_{\mathbf{3}}
~\otimes~
\left(
\begin{array}{c}
\beta_1
\\
\beta_2
\\
\beta_3
\end{array}
\right)_{\mathbf{\overline{3}}}
\nonumber\\
&=&~~
(\alpha_1 \beta_1 + \alpha_2 \beta_2 + \alpha_3 \beta_3)_{ \mathbf{1}_{0,0}}
\oplus
( \alpha_1 \beta_1 + \omega^2 \alpha_2 \beta_2 + \omega \alpha_3 \beta_3)_{ \mathbf{1}_{1,0}}
\oplus
( \alpha_1 \beta_1 +  \omega \alpha_2 \beta_2 + \omega^2 \alpha_3 \beta_3)_{ \mathbf{1}_{2,0}}
\nonumber\\
&&\oplus
( \alpha_3 \beta_1 + \alpha_1 \beta_2 + \alpha_2 \beta_3)_{ \mathbf{1}_{0,1}}
\oplus
( \alpha_3 \beta_1 + \omega^2 \alpha_1 \beta_2 + \omega \alpha_2 \beta_3)_{ \mathbf{1}_{1,1}}
\oplus
( \alpha_3 \beta_1 +  \omega \alpha_1 \beta_2 + \omega^2 \alpha_2 \beta_3)_{ \mathbf{1}_{2,1}}
\nonumber\\
&&\oplus
( \alpha_2 \beta_1 + \alpha_3 \beta_2 + \alpha_1 \beta_3)_{ \mathbf{1}_{0,2}}
\oplus
( \alpha_2 \beta_1 + \omega^2 \alpha_3 \beta_2 + \omega \alpha_1 \beta_3)_{ \mathbf{1}_{1,2}}
\oplus
( \alpha_2 \beta_1 +  \omega \alpha_3 \beta_2 + \omega^2 \alpha_1 \beta_3)_{ \mathbf{1}_{2,2}}\,.
\end{eqnarray*}

\item{ $\mathbf{3}\otimes\mathbf{3}= \mathbf{\overline{3}}_{S_1} \oplus   \mathbf{\overline{3}}_{S_2} \oplus \mathbf{\overline{3}}_A$}
\begin{eqnarray}
\hskip-0.5cm\left(
\begin{array}{c}
\alpha_1
\\
\alpha_2
\\
\alpha_3
\end{array}
\right)_{\mathbf{3}}
~\otimes&
\left(
\begin{array}{c}
\beta_1
\\
\beta_2
\\
\beta_3
\end{array}
\right)_{\mathbf{3}}
&=~
\left(
\begin{array}{c}
\alpha_1 \beta_1
\\
\alpha_2 \beta_2
\\
\alpha_3 \beta_3
\end{array}
\right)_{\mathbf{\overline{3}}_{S_1}}
\oplus
\frac{1}{2}
\left(
\begin{array}{c}
\alpha_2 \beta_3 + \alpha_3 \beta_2
\\
\alpha_3 \beta_1 + \alpha_1 \beta_3
\\
\alpha_1 \beta_2 + \alpha_2 \beta_1
\end{array}
\right)_{\mathbf{\overline{3}}_{S_2}}
\oplus
\frac{1}{2}
\left(
\begin{array}{c}
\alpha_2 \beta_3 - \alpha_3 \beta_2
\\
\alpha_3 \beta_1 - \alpha_1 \beta_3
\\
\alpha_1 \beta_2 - \alpha_2 \beta_1
\end{array}
\right)_{\mathbf{\overline{3}}_A}\,,
\end{eqnarray}
where the subscripts ``\textit{S}'' and ``\textit{A}'' denote symmetric and anti-symmetric combinations respectively.

\item{$\mathbf{\overline{3}}\otimes\mathbf{\overline{3}}=\mathbf{3}_{S_1}      \oplus\mathbf{3}_{S_2}\oplus\mathbf{3}_A$}
\begin{eqnarray}
\hskip-0.5cm\left(
\begin{array}{c}
\alpha_1
\\
\alpha_2
\\
\alpha_3
\end{array}
\right)_{\mathbf{\overline{3}}}
~\otimes&
\left(
\begin{array}{c}
\beta_1
\\
\beta_2
\\
\beta_3
\end{array}
\right)_{\mathbf{\overline{3}}}
&=~
\left(
\begin{array}{c}
\alpha_1 \beta_1
\\
\alpha_2 \beta_2
\\
\alpha_3 \beta_3
\end{array}
\right)_{\mathbf{3}_{S_1}}
\oplus
\frac{1}{2}
\left(
\begin{array}{c}
\alpha_2 \beta_3 + \alpha_3 \beta_2
\\
\alpha_3 \beta_1 + \alpha_1 \beta_3
\\
\alpha_1 \beta_2 + \alpha_2 \beta_1
\end{array}
\right)_{\mathbf{3}_{S_2}}
\oplus
\frac{1}{2}
\left(
\begin{array}{c}
\alpha_2 \beta_3 - \alpha_3 \beta_2
\\
\alpha_3 \beta_1 - \alpha_1 \beta_3
\\
\alpha_1 \beta_2 - \alpha_2 \beta_1
\end{array}
\right)_{\mathbf{3}_A}\,.
\end{eqnarray}

\end{itemize}

\section{\label{sec:vacuum_alignment}Vacuum Alignment}

In this Appendix, we shall investigate the problem of achieving the
vacuum configuration in Eq.~\eqref{Cases} and
Eq.~\eqref{FlavonVEV}. For self-consistency, all flavon fields
$\varphi$, $\xi$, $\sigma_1$ and $\sigma_2$ are treated as complex,
given the form of the $\Delta(27)$ representation matrices, and the
fact that the $Z_4$ charge of $\sigma_2$ is purely imaginary. Since
the flavons $\varphi$ and $\xi$, $\sigma_1$, $\sigma_2$ are assumed to
be localized at $y=0$ and $y=L$ respectively, the vacuum alignment
problem is greatly simplified.
At the UV brane $y=0$, the flavon $\varphi$ transforms in the manner
listed in Table~\ref{tl}. The scalar potential invariant under the
flavor symmetry $\Delta(27)\otimes Z_4\otimes Z^{\prime}_4$ can be
written as:
\begin{eqnarray}
V_{\text{UV}} &= &M_{\varphi}^2(\varphi\varphi^{\ast})_{\mathbf{1}_{0,0}}+f_1\left[(\varphi\varphi)_{\mathbf{\overline{3}}_{S_1}} (\varphi^{\ast}\varphi^{\ast})_{\mathbf{3}_{S_1}}\right]_{\mathbf{1}_{0,0}}
+f_2\left[(\varphi\varphi)_{\mathbf{\overline{3}}_{S_2}}(\varphi^{\ast}\varphi^{\ast})_{\mathbf{3}_{S_2}}\right]_{\mathbf{1}_{0,0}}\nn\\ &&+f_3\left[(\varphi\varphi)_{\mathbf{\overline{3}}_{S_1}}(\varphi^{\ast}\varphi^{\ast})_{\mathbf{3}_{S_2}}\right]_{\mathbf{1}_{0,0}}
+f_3^{\ast}\left[(\varphi\varphi)_{\mathbf{\overline{3}}_{S_2}}(\varphi^{\ast}\varphi^{\ast})_{\mathbf{3}_{S_1}}\right]_{\mathbf{1}_{0,0}}\,,
\end{eqnarray}
with real couplings $M_{\varphi}^2$, $f_1$ and $f_2$. Note that
$\varphi=(\varphi_1, \varphi_2, \varphi_3)$ is a $\Delta(27)$ triplet
$\mathbf{3}$, and its complex conjugate
$\varphi^{\ast}=(\varphi^{\ast}_1, \varphi^{\ast}_2,
\varphi^{\ast}_3)$ transforms consequently as $\mathbf{\overline{3}}$.
Focusing on the field configuration
\begin{eqnarray}
\langle \varphi \rangle=(1,1,1) v_\varphi\,,
\end{eqnarray}
the minimum conditions for the UV potential read
\begin{eqnarray}
\frac{\partial V_{\text{UV}}}{\partial \varphi_1^*}=
\frac{\partial V_{\text{UV}}}{\partial \varphi_2^*}=
\frac{\partial V_{\text{UV}}}{\partial \varphi_3^*}=
v_\varphi\left[M_\varphi^2
+ 2(f_1 + f_2 +f_3 + f_3^*)  \big|v_\varphi\big|^2\right]
=0\,,
\end{eqnarray}
leading to a non zero solution
\begin{eqnarray}
\big|v_\varphi\big|^2=- \frac{ M_\varphi^2 }{ 2(f_1 + f_2 + f_3 + f_3^*) }\,,
\end{eqnarray}
that holds in a finite portion of parameter space with $f_1 + f_2 +
f_3 + f_3^*<0$.

Similarly, at the IR brane $y=L$, the most general renormalizable
scalar potential $V_{\text{IR}}$ involving the flavon fields $\xi$,
$\sigma_1$, $\sigma_2$ is
\begin{eqnarray}
V_{\text{IR}} &=&
M_{\xi}^2(\xi\xi^{\ast})_{\mathbf{1}_{0,0}}+M_{\sigma_1}^2(\sigma_1\sigma^{\ast}_1)_{\mathbf{1}_{0,0}}+ M_{\sigma_2}^2(\sigma_2\sigma^{\ast}_2)_{\mathbf{1}_{0,0}}+g_1\left[(\xi\xi)_{\mathbf{\overline{3}}_{S_1}}(\xi^{\ast}\xi^{\ast})_{\mathbf{3}_{S_1}}\right]_{\mathbf{1}_{0,0}}\nn\\
&&+g_2\left[(\xi\xi)_{\mathbf{\overline{3}}_{S_2}}(\xi^{\ast}\xi^{\ast})_{\mathbf{3}_{S_2}}\right]_{\mathbf{1}_{0,0}}+g_3\left[(\xi\xi)_{\mathbf{\overline{3}}_{S_1}}(\xi^{\ast}\xi^{\ast})_{\mathbf{3}_{S_2}}\right]_{\mathbf{1}_{0,0}}+ g^{\ast}_3\left[(\xi\xi)_{\mathbf{\overline{3}}_{S_2}}(\xi^{\ast}\xi^{\ast})_{\mathbf{3}_{S_1}}\right]_{\mathbf{1}_{0,0}}\nn\\ &&+g_4\sigma_1^2\sigma_1^{\ast2}+g_5\sigma_2^2\sigma_2^{\ast2}
+g_6|\sigma_1|^2|\sigma_2|^2+g_7(\xi\xi^{\ast})_{\mathbf{1}_{0,0}}|\sigma_1|^2+g_8(\xi\xi^{\ast})_{\mathbf{1}_{0,0}}|\sigma_2|^2\nn\\ &&+g_9(\xi\xi^{\ast})_{\mathbf{1}_{0,1}}\sigma_1^2+ g^{\ast}_9(\xi\xi^{\ast})_{\mathbf{1}_{0,2}}\sigma^{\ast2}_1\,,
\end{eqnarray}
where all couplings, excluding $g_3$ and $g_9$, are real. For this potential, the Case I alignment
\begin{eqnarray}
\langle \xi \rangle=(0,v_\xi,0)\,,
\hspace{0.6cm}
\langle \sigma_1 \rangle=v_{\sigma_1}\,,
\hspace{0.6cm}
\langle \sigma_2 \rangle=v_{\sigma_2}\,,
\end{eqnarray}
determines the minimization conditions
\begin{eqnarray}
\frac{\partial V_\text{IR}}{\partial \xi_1^*}  &=&
g_9^* v_\xi {v_{\sigma_1}^*}^2
=0\,,
\nn\\
\frac{\partial V_\text{IR}}{\partial \xi_2^*}&=&
v_\xi\Big(M_\xi^2+2 g_1 \big|v_\xi\big|^2+g_7\big|v_{\sigma_1}\big|^2+g_8\big|v_{\sigma_2}\big|^2\Big)
=0\,,
\nn\\
\frac{\partial V_\text{IR}}{\partial \xi_3^*}  &=&
g_9 v_\xi v_{\sigma_1}^2
=0\,,
\nn\\
\frac{\partial V_\text{IR}}{\partial\sigma_1^*}&=&
v_{\sigma_1}\Big(M_{\sigma_1}^2+2g_4\big|v_{\sigma_1}\big|^2+g_7\big|v_\xi\big|^2+g_6\big|v_{\sigma_2}\big|^2\Big)
=0\,,
\nn\\
\frac{\partial V_\text{IR}}{\partial\sigma_2^*}&=&
v_{\sigma_2}\Big(M_{\sigma_2}^2+2g_5\big|v_{\sigma_2}\big|^2+g_8\big|v_\xi\big|^2+g_6\big|v_{\sigma_1}\big|^2\Big)
=0\,.
\end{eqnarray}
From the above equations, it is clear that non-trivial solutions in this sector are only achievable by fine tuning the $g_9$ parameter to satisfy $g_9= 0$. This choice can be enforced by an additional dynamical mechanism capable of switching off the $(\xi\xi^{\ast})_{\mathbf{1}_{0,1}}\sigma_1^2$ and $(\xi\xi^{\ast})_{\mathbf{1}_{0,2}}\sigma^{\ast2}_1$ terms in the potential. Such scenario could be naturally realized in a supersymmetric extension ~\cite{Altarelli:2005yp,Altarelli:2005yx}. As this possibility lies beyond the scope or the present work, we simply impose the condition $g_9= 0$ in the general potential.  Then, the obtained solutions are given by
\begin{eqnarray}
|v_\xi|^2&=&\frac{(g_6^2-4g_4g_5)M_{\xi}^2+(2g_5g_7-g_6g_8)M_{\sigma_1}^2+(2g_4g_8-g_6g_7)M_{\sigma_2}^2}{2(4g_1g_4g_5+g_6g_7g_8-g_1g_6^2-g_4g_8^2-g_5g_7^2)}\,,\nn\\
|v_{\sigma_1}|^2&=&\frac{(2g_5g_7-g_6g_8)M_{\xi}^2+(g_8^2-4g_1g_5)M_{\sigma_1}^2+(2g_1g_6-g_7g_8)M_{\sigma_2}^2}{2(4g_1g_4g_5+g_6g_7g_8-g_1g_6^2-g_4g_8^2-g_5g_7^2)}\,,\nn\\
|v_{\sigma_2}|^2&=&\frac{(2g_4g_8-g_6g_7)M_{\xi}^2+(2g_1g_6-g_7g_8)M_{\sigma_1}^2+(g_7^2-4g_1g_4)M_{\sigma_2}^2}{2(4g_1g_4g_5+g_6g_7g_8-g_1g_6^2-g_4g_8^2-g_5g_7^2)}\,.
\end{eqnarray}
The right-handed side of these expressions can be positive in a finite region of parameter space.
Analogously, for the Case II vacuum configuration
\begin{eqnarray}
\langle \xi \rangle=(1,\omega,1) v_\xi\,,
\hspace{0.6cm}
\langle \sigma_1 \rangle=v_{\sigma_1}\,,
\hspace{0.6cm}
\langle \sigma_2 \rangle=v_{\sigma_2}\,,
\end{eqnarray}
the minimization conditions are
\begin{eqnarray}
\frac{\partial V_\text{IR}}{\partial\xi^{\ast}_1}&=&v_{\xi}\Big[    M_\xi^2+2\big(g_1+g_2+\omega^2g_3+\omega g^{\ast}_3\big)\big|v_\xi\big|^2+g_7\big|v_{\sigma_1}\big|^2
+g_8\big|v_{\sigma_2}\big|^2+g_9v_{\sigma_1}^2+\omega g^{\ast}_9v^{\ast2}_{\sigma_1}\Big]=0,\nn\\
\frac{\partial V_\text{IR}}{\partial\xi^{\ast}_2}&=&\omega v_{\xi}\Big[    M_{\xi}^2+2\big(g_1+g_2+\omega^2 g_3+\omega g^{\ast}_3\big)\big|v_\xi\big|^2+g_7\big|v_{\sigma_1}\big|^2+g_8\big|v_{\sigma_2}\big|^2 +\omega^2g_9v_{\sigma_1}^2+\omega^2g^{\ast}_9v^{\ast2}_{\sigma_1}\Big]=0,\nn\\
\frac{\partial V_\text{IR}}{\partial \xi_3^{\ast}}&=&v_{\xi}\Big[M_\xi^2 +2\big(g_1+g_2+\omega^2 g_3+\omega g_3^{\ast}\big)\big|v_{\xi}\big|^2+g_7\big|v_{\sigma_1}\big|^2+g_8\big|v_{\sigma_2}\big|^2 +\omega g_9v_{\sigma_1}^2+g_9^{\ast}v^{\ast2}_{\sigma_1}\Big]=0,\nn\\
\frac{\partial V_\text{IR}}{\partial\sigma_1^{\ast}}&=&v_{\sigma_1}\Big[ M_{\sigma_1}^2+2g_4\big|v_{\sigma_1}\big|^2+g_6\big|v_{\sigma_2}\big|^2+3g_7\big|v_{\xi}\big|^2\Big]=0,\nn\\
\frac{\partial V_\text{IR}}{\partial\sigma_2^{\ast}}&=&v_{\sigma_2}\Big[M_{\sigma_2}^2+2g_5\big|v_{\sigma_2}\big|^2+g_6\big|v_{\sigma_1}\big|^2+3g_8\big|v_\xi\big|^2\Big]=0\,.
\end{eqnarray}
Again, these equations are incompatible unless $g_{9}=0$. Once the
coupling $g_{9}$ is enforced to vanish, we are left with three
independent linear equations for the three unknown variables
$v_{\xi}$, $v_{\sigma_1}$ and $v_{\sigma_2}$. The solutions can be
easily found as
\begin{eqnarray}
|v_\xi|^2&=&\frac{(4g_4g_5-g_6^2)M_{\xi}^2+(g_6g_8-2g_5g_7)M_{\sigma_1}^2+(g_6g_7-2g_4g_8)M_{\sigma_2}^2}{2\widetilde{g}(g_6^2-4g_4g_5)+6(g_4g_8^2+g_5g_7^2-g_6g_7g_8)}\,,\nn\\
|v_{\sigma_1}|^2&=&\frac{(3g_6g_8-6g_5g_7)M_{\xi}^2+(4\widetilde{g}g_5-3g_8^2)M_{\sigma_1}^2+(3g_7g_8-2\widetilde{g}g_6)M_{\sigma_2}^2}{2\widetilde{g}(g_6^2-4g_4g_5)+6(g_4 g_8^2+g_5g_7^2-g_6g_7g_8)}\,,\nn\\
|v_{\sigma_2}|^2&=&\frac{(3g_6g_7-6g_4g_8)M_{\xi}^2+(3g_7g_8-2\widetilde{g}g_6)M_{\sigma_1}^2+(4\widetilde{g}g_4-3g_7^2)M_{\sigma_2}^2}{2\widetilde{g}(g_6^2-4g_4g_5)+6(g_4 g_8^2+g_5g_7^2-g_6g_7g_8)}\,.
\end{eqnarray}
where we have defined $\widetilde{g}\equiv g_1+g_2 +\omega^2g_3+\omega
g_3^{\ast}$. Therefore, both $\langle\xi\rangle=(0, v_\xi, 0)$ and
$\langle\xi\rangle=(1, \omega, 1)v_\xi$ alignments can describe the
local minimum of $V_{\text{IR}}$, depending on the the parameter
values. In the case of $g_{1}\ll\widetilde{g}$, the VEV
$\langle\xi\rangle=(0, v_\xi, 0)$ is preferred over
$\langle\xi\rangle=(1, \omega, 1)v_\xi$, while $V_{\text{IR}}$ is
minimized by $\langle\xi\rangle=(1, \omega, 1)v_\xi$ for
$g_{1}\gg\widetilde{g}$.

\end{appendix}


\end{document}